\documentclass[]{aa}  
\usepackage{amsmath}
\usepackage{graphicx}
\usepackage{txfonts}
\usepackage{natbib} 
\bibpunct{(}{)}{;}{a}{}{,} % to follow the A&A style
 
\usepackage{verbatim}
\usepackage{url}
\usepackage{epstopdf}

\usepackage{color}

\begin{document}

   \title{Relevance of the small frequency separation for asteroseismic stellar age, mass, and radius} 

   \subtitle{A statistical investigation for main-sequence low-mass stars}

   \author{G. Valle \inst{1,2}, M. Dell'Omodarme \inst{1}, P.G. Prada Moroni
     \inst{1,2}, S. Degl'Innocenti \inst{1,2} 
          }
   \titlerunning{Asteroseismic estimates of fundamental parameters}
   \authorrunning{Valle, G. et al.}

   \institute{
Dipartimento di Fisica "Enrico Fermi'',
Universit\`a di Pisa, Largo Pontecorvo 3, I-56127, Pisa, Italy
\and
 INFN,
 Sezione di Pisa, Largo Pontecorvo 3, I-56127, Pisa, Italy
 }

   \offprints{G. Valle, valle@df.unipi.it}

   \date{Received 22/07/2019; accepted 03/02/2020}

  \abstract
{}
{We performed a theoretical analysis aimed at quantifying the relevance of the small frequency separation $\delta \nu$ in determining stellar ages, masses, and radii. We aimed to establish a minimum uncertainty on these quantities for low-mass stars across different evolutionary stages of the main sequence and to evaluate the biases that come from some systematic differences between the stellar model grid adopted for the recovery and the observed stars. }
{We adopted the SCEPtER (Stellar CharactEristics Pisa
Estimation gRid) pipeline for low-mass stars, [0.7, 1.05] $M_{\sun}$, from the zero-age main sequence to the central hydrogen depletion. For each model in the grid, we computed oscillation frequencies. Synthetic stars were generated and reconstructed based on different assumptions about the relative precision in the $\delta \nu$ parameter (namely 5\% and 2\%).
The quantification of the systematic errors arising from a possible mismatch between synthetic stars and the recovery grid was performed by generating stars from synthetic grids of stellar models with different initial helium abundance and microscopic diffusion efficiency. The results obtained without $\delta \nu$ as an observable are included for comparison.}
{The investigation highlighted and confirmed the improvement in the age estimates when $\delta \nu$ is available, which has already been reported in the literature. While the biases were negligible, the statistical error affecting age estimates was strongly dependent on the stellar evolutionary phase. The error is at its maximum at ZAMS and it decreases to about 11\% and 6\% ($\delta \nu$ known at 5\% and 2\% level, respectively) when stars reach the 30\% of their evolutionary MS lifetime. The usefulness of small frequency separation in improving age estimates vanishes in the last 20\% of the MS. 
The availability of $\delta \nu$ in the fit for mass and radius estimates provided an effect that was nearly identical to its effect on age, assuming an observational uncertainty of 5\%. 
As a departure, with respect to age estimates, no benefit was detected for mass and radius determinations from a reduction of the observational error in $\delta \nu$ to 2\%.         
The age variability attributed to differences in the initial helium abundance resulted in negligible results owing to compensation effects that have already been discussed in previous works. 
On the other hand, the current uncertainty in the initial helium abundance leads to a greater bias (2\% and 1\% level) in mass and radius estimates whenever $\delta \nu$ is in the observational pool. This result, together with the presence of further unexplored uncertainty sources, suggest that precision in the derived stellar quantities below these thresholds may possibly be overoptimistic.   
The impact of microscopic diffusion was investigated by adopting  a grid of models for the
recovery which totally neglected the process. The availability of the small frequency separation resulted in biases lower than 5\% and 2\% for observational errors of 5\% and 2\%, respectively.
The estimates of mass and radius showed again a greater distortion when $\delta \nu$ is included among the observables. These biases are at the level of 1\%, confirming that threshold as a minimum realistic uncertainty on the derived stellar quantities.  
Finally, we compared the estimates by the SCEPtER pipeline for 13 {\it Kepler} asteroseismic LEGACY sample stars with those given by six different pipelines from literature. This procedure demonstrated a fair agreement for the results. The comparison suggests that a realistic approach to the determination of the error on the estimated parameters consists of approximately doubling the error in the recovered stellar characteristics from a single pipeline. Overall, on the LEGACY sample data, we obtained a multi-pipeline precision of about 4.4\%, 1.7\%, and 11\% on the estimated masses, radii, and ages, respectively. }
{}


   \maketitle

\section{Introduction}\label{sec:intro}

Asteroseismology allows us to constrain stellar masses, radius, and age with unprecedented precision.
Global asteroseismic parameters, namely the large frequency separation $\Delta \nu$ and the frequency of the maximum oscillation  power $\nu_{\rm max}$, are  routinely used  today to
determine stellar properties \citep[see  e.g.][]{Basu2010, Quirion2010, Chaplin2011, Silva2012,  Chaplin2013, Chaplin2014, eta, bulge, Serenelli2017,Rodrigues2017, Pinsonneault2018}.  While stellar mass and radius estimates are known to be reliable  at the level of a few percent when $\Delta \nu$ and $\nu_{\rm max}$ are known, the ages are much less constrained. In fact, while these two asteroseismic quantities have a strong dependency on mass and radius, they are 
not particularly sensitive to the deep stellar layers, which change most dramatically during the main sequence (MS) evolution \citep[see e.g.][]{Aerts2010}.

The  increasing availability of oscillation frequencies for target stars in recent years has been coupled with an increasing precision with regard to their determinations, carrying with it an obvious impact on the precision of stellar age determinations. The evaluation of $\Delta \nu$ and $\nu_{\rm max}$ at some percent level was common few years ago \citep[e.g.][]{Mathur2012, Chaplin2014} but much better results can be achieved today thanks to longer observation periods. 
Currently, the set of the best asteroseismically characterised MS stars -- the {\it Kepler} asteroseismic LEGACY sample \citep{Lund2017} -- contains 66 stars with an exceptionally high signal-to-noise ratio (S/N), with median precision in global asteroseismic parameters of around 0.05\% in $\Delta \nu$ and 0.6\% in $\nu_{\rm max}$.

Besides these global parameters, which can be extracted from oscillation spectra for most of the targets, there is a growing amount of stars with precise measurements of individual frequencies that can be directly fitted \citep[e.g.][]{Eggenberger2004, Metcalfe2009, Howell2012, Metcalfe2012, Metcalfe2014, Metcalfe2015, Campante2015,SilvaAguirre2015, SilvaAguirre2017, Bellinger2016, Aerts2018}. 
Most of the studies in the field adopt a grid-based maximum-likelihood or a Bayesian approach in the fitting, in addition to emerging machine-learning methods  \citep[see e.g.][]{Bellinger2016, Aerts2018,Hendriks2018}. 

Even when individual frequencies are not directly used, their availability can be exploited to compute other global seismic parameters, in particular the so called small frequency  separation $\delta \nu$, as defined in Sect.~\ref{sec:fittingML} \citep{Dalsgaard1993}.
The great improvement that come from the knowledge of individual frequency or $\delta \nu$ concerns the age estimates, with a typical improvement in the precision by a factor of two 
\citep[see e.g.][]{Miglio2005, Mathur2012, Lebreton2014, SilvaAguirre2015}. The recent literature also includes other approaches to define seismic indicators to better constrain stellar parameters \citep[see e.g.][]{Farnir2019}.

Although several studies have addressed specific questions about the relevance of $\delta \nu$, or of the frequency ratios defined by \citet{Roxburgh2003}, in the stellar characteristics estimates
\citep[e.g.][in the recent literature]{Brandao2013, Deheuvels2015, Bellinger2016, Angelou2017, Aerts2018, Rendle2019}, some theoretical problems  still remain unexplored; in particular, the quantification of the minimum uncertainty as a function of different assumptions on the observational errors in the estimates of stellar mass, radius, and age across different evolutionary stages. A direct quantification will help to understand the presence of possible hidden biases that can affect the estimated properties.

Another open question concerns the evaluation of the biases that come from some systematic differences between the stellar models grid adopted for the recovery and real world stars.
Some studies in the literature  \citep[e.g.][]{SilvaAguirre2015,Nsamba2018} address these problems on a sample of real-world stars by adopting different  evolution  and  pulsation  codes into the fit. However, these analyses cannot control the distortion owing to the unknown systematic differences between the recovery grids and the evolution of real world stars. A firmer theoretical grasp of these questions can only be gained through controlled simulations.  

Several works explored these aspects when only $\Delta \nu$ and $\nu_{\rm max}$ were available \citep[e.g.][]{Gai2011,basu2012,scepter1,eta}. The aim of this paper is to propose an analysis similar to that in \citet{eta}, but while considering $\delta \nu$  in the observational pool.

\section{Methods}\label{sec:method}

With the aim of investigating the relevance of the small frequency separation constraint for the estimation of stellar characteristic, we adopted a modified SCEPtER (Stellar CharactEristics Pisa
Estimation gRid) pipeline \citep{scepter1,eta,bulge}. We performed the analysis  based on a specific range of stellar models in the main-sequence (MS) and for low-mass stars (below 1.05 $M_{\sun}$). This choice allowed us to neglect the uncertainty in the convective core overshooting extension since the stars in the chosen mass range burn hydrogen in a radiative core. As a matter of fact the extension of the present investigation to an higher mass range will be affected not only by this supplementary degree of freedom, which would force the building of a huge dataset of stellar tracks, but also by the requirement of following the evolution of the convective core with a spatial resolution much higher than the one needed for the present investigation. Moreover, \citet{Deal2018} point out that more massive stars may suffer from additional biases because of the effect of radiative accelerations.
   
Section~\ref{sec:grids} describes in detail the grid of stellar models adopted in the computations, while Sect.~\ref{sec:freq} reports the details of the frequency computation from stellar models.  
Finally, Sect.~\ref{sec:fittingML} describes the estimation process based on a series of Monte Carlo experiments.

\subsection{Stellar model grid}
\label{sec:grids}

The model grids were computed for  masses in the range [0.7, 1.05] $M_{\sun}$, with a step of 0.01 $M_{\sun}$. The evolution was followed from the pre-MS until the exhaustion of the central hydrogen, conventionally assumed when its central abundances drops below $10^{-5}$, which corresponds to a star in the early sub-giant branch. 
Only models in MS and with age lower than 14 Gyr were retained in the grid.
The initial metallicity [Fe/H] was varied from $-0.5$ to 0.3 dex, with
a step of 0.05 dex. 
The solar heavy-element mixture by \citet{AGSS09} was adopted. 
Five initial helium abundances were considered at fixed metallicity by adopting the commonly used
linear relation $Y = Y_p+\frac{\Delta Y}{\Delta Z} Z$
with the primordial abundance $Y_p = 0.2485$ from WMAP
\citep{peimbert07a,peimbert07b} and with a helium-to-metal enrichment ratio $\Delta Y/\Delta Z$
from 1 to 3 with a step of 0.5 \citep{gennaro10}. We adopted $\Delta Y/\Delta Z = 2.0$ in the reference scenario and used other grids to assess the importance of a systematic mismatch between synthetic data and recovery grid.

Nuclear reaction rates were taken from the NACRE compilation \citep{nacre},
except for $^{14}$N$(p,\gamma)^{15}$O; for this, we adopted the estimates of \citet{14n}.
Microscopic diffusion was considered according to \citet{thoul94}, with radiation turbulence by \citet{Morel2002}, as adopted in \citet{Choi2016}. A grid that neglects microscopic diffusion was also computed and used to test the importance of this input physics in the final estimates. 
A solar-calibrated mixing-length parameter $\alpha_{\rm ml} = 1.8$ was adopted.  Outer boundary conditions were set by the
atmospheric models from \citet{hauschildt99}, supplemented with models by \citet{castelli03}, where the former were unavailable. 
Convective core overshooting was not included because the stars in the mass range adopted for the present analysis have a radiative core.

The MESA stellar evolutionary code \citep[r10398;][]{MESA2011,MESA2013,MESA2015, MESA2018} was used for the computations. To obtain sensible oscillation frequencies, about  2\,000 meshes were placed in the stellar structures \citep{Moya2008}. As a control, we also built a restricted dataset with about 800 meshes per structure. The comparison of the results obtained with the two grids showed a variation in the estimates that were much smaller than the statistical errors from observational uncertainty propagation. 

The adopted configuration closely matches the one used by \citet{scepter1, eta}, which allows for an easy comparison of the results. The quoted researches used the FRANEC code \citep{scilla2008, Tognelli2011}, in the configuration adopted to compute the Pisa Stellar
Evolution Data Base\footnote{\url{http://astro.df.unipi.it/stellar-models/}} 
for low-mass stars \citep{database2012}. As we directly verified in the past \citep{testW, TZFor},
the difference in the results between FRANEC and MESA stellar evolutionary codes, adopted in the same configuration, are marginal.

\subsection{Frequencies computation}
\label{sec:freq}

Oscillation frequencies for each stellar model, from ZAMS to the central hydrogen exhaustion, 
were calculated using GYRE \citep{Townsend2013, Townsend2018}. We adopted it in the standard configuration, solving the 4th order system of pulsation equations. A static atmosphere model and the standard inner
and outer boundary conditions was adopted, as outlined in Appendix A of \citet{Townsend2013}. We considered these in computations radial modes (degree $l$ = 0) and non-radial modes of degree $l$ = 1, 2.

Let us denote as $S$ the difference between a frequency $\nu$ of spherical degree $l_1$ and radial order $n_1$ and a frequency of $l_2$ and $n_2$,
\begin{equation}
S(n_1, n_2, l_1, l_2) \equiv  \nu_{l_1}(n_1) - \nu_{l_2}(n_2).\label{eq:diff-freq}
\end{equation}
The large frequency separation is then given by
\begin{equation}
\Delta \nu_l(n) \equiv S(n, n-1, l, l),\label{eq:Dnu}
\end{equation}
while the small frequency separation is
\begin{equation}
\delta \nu_{(l,l+2)}(n) \equiv S(n, n-1, l, l+2).\label{eq:dnu}
\end{equation}
Values for
 $\Delta \nu_0$ and $\delta \nu_{02}$ were computed for each stellar model in the grid. The observed oscillation power spectrum of solar-like stars is characterised by a typical Gaussian-like envelope, peaked at $\nu_{\rm max}$. Therefore,
to mimic the observable frequency ranges for each model, the frequencies were weighted considering their position in a Gaussian envelope. This envelope was centred at the predicted frequency of maximum oscillation
power, given by scaling relation \citep{Kjeldsen1995}:
\begin{equation}\label{eq:nimax}
\frac{\nu_{\rm
                max}}{\nu_{\rm max, \sun}} =  \frac{{M/M_{\sun}}}{ (R/R_{\sun})^2
        \sqrt{ T_{\rm eff}/T_{\rm eff, \sun}} }. 
\end{equation}
The full-width at half-maximum of the envelope was taken as $0.66 \, \nu_{\rm max}^{0.88}$ \citep{Mosser2012, Bellinger2016}. The weighted median over $n$, $\langle \Delta \nu_0 \rangle$ and $\langle \delta \nu_{02} \rangle$,
where then computed and adopted as asteroseismic observational constraints \citep{Bellinger2016}, along with $\nu_{\rm max}$\footnote{The computations relied on a customised version of the scripts provided by E. Bellinger \url{https://github.com/earlbellinger/asteroseismology}.}.
For this last parameter, it worth noting that the validity of the scaling relations (mainly in the RGB phase) has been questioned in recent years \citep{Epstein2014, Gaulme2016, Viani2017, Brogaard2018, Buldgen2019}, posing a serious problem  whenever it is adopted for a comparison with observational data. 
Moreover, it is well known that for stars with convective envelopes such as those considered in the present research, the treatment of convection by 1D schemes (as the mixing-length theory adopted here, \citealt{bohmvitense58}) leads to inaccurate representations of the surface layers, which, in turn, has an impact on the predicted stellar frequencies. The resulting systematic discrepancies between computed and observed frequencies is usually referred to as the surface effect \citep[see e.g.][]{Brown1984,Dalsgaard1988}. Fortunately,  both these issues are of minor importance for our aims, because both artificial data and the recovery grid are computed using the same scheme.
For ease in notation, in the following, we refer to $\langle \Delta \nu_0 \rangle$ and $\langle \delta \nu_{02} \rangle$ simply as $\Delta \nu$ and $\delta \nu$.

\subsection{Fitting procedures}\label{sec:fittingML}

The fit was performed by means of the SCEPtER pipeline \citep{eta, bulge, binary}, a well-tested technique that has been adopted in the past for single and binary stars. We briefly summarise the technique here for reader's convenience. 

For each point $j$ on the grid, we define $q \equiv \{T_{\rm eff}, {\rm [Fe/H]}, \Delta \nu, \delta \nu, \nu_{\rm max}\}$ as the vector of observed quantities for a star, and $\sigma$ to be the vector of the corresponding observational uncertainties. We also define $\tilde q_j$ as the vector of observables for each $j$th point of the grid.
We compute the geometrical distance $d_{j}$ between the observed star and the $j$th grid point, defined as:
\begin{equation}
d_{j} = \left\lVert \frac{q - \tilde q_j}{\sigma} \right\rVert. \label{eq:dist}
\end{equation} 
Then the technique computes the likelihood as:
\begin{equation}
L_j = \exp (-d_{j}^2/2).
\label{eq:lik-scepter}
\end{equation}

This likelihood function is evaluated for each grid point within $3 \sigma$ of
all the variables from $q$; we define $L_{\rm max}$ as the maximum value
obtained in this step. The estimated stellar quantities are obtained
by averaging the corresponding mass, radii, and age of all the models with a likelihood
greater than $0.95 \times L_{\rm max}$.

This technique can also be employed to construct a Monte Carlo confidence
interval for mass, radius, and age estimates. Starting from observational values $q$ and their uncertainties $\sigma$, a synthetic sample of $n$ stars is
generated, following a multivariate normal distribution with vector of mean
$q$ and covariance matrix $\Sigma = {\rm diag}(\sigma)$. A value of
$n = 5\,000$ can usually provide a fair balance between
computation time and the accuracy of the results.  The medians of the $n$
objects mass, radius, and age are taken as the best estimate of the true values;
the 16th and 84th quantiles of the $n$ values are adopted as a $1 \sigma$
confidence interval. 

\section{Estimated parameters}
\label{sec:results}

We aim to quantify the biases and the propagation of errors from the observational constraints to the final estimates of stellar age, mass, and radius. In particular, we are interested in the analysis of how these quantities change as a star evolves through the whole MS and if there are some effects connected to the mass or initial metallicity \citep[see e.g.][for a recent analysis of statistical distorsions that can affect the mixing-length parameter estimate for field stars]{ML}. We performed these evaluations relying on synthetic stars sampled from a grid of stellar models. We  explored both ideal scenarios -- in which the grid and the mock data are in perfect agreement beside a random perturbation that accounts for observational uncertainties -- and configurations that assume systematic discrepancies between data and recovery grid (Sect.~\ref{sec:syst}).    

As a starting point, we verified the performance of the best possible configuration for adopting the same stellar models for the mock data generation and for the estimation procedure. 
Therefore, we built a synthetic dataset
by sampling $N = 50\,000$ artificial stars from the same standard estimation
grid of stellar models used in the recovery procedure ($\Delta Y/\Delta Z = 2.0$), adding to each of them
a Gaussian noise in all the observed quantities. We assumed  0.5\% in $\Delta \nu$, 0.7\% in
$\nu_{\rm max}$, 5\% in $\delta \nu$, 100 K in $T_{\rm eff}$, and 0.1 dex in [Fe/H].
The errors in the asteroseismic quantities were set by considering the quoted uncertainties in the {\it Kepler} asteroseismic LEGACY sample \citep{Lund2017} as the absolute maximum reachable precision, and increasing the errors by about one order of magnitude. The adopted uncertainties are expected to be attainable for a large share of stars. A second scenario considers an optimistic error in $\delta \nu$ of 2\%, which is the median error in this parameter for LEGACY sample stars.
  
This Monte Carlo simulation proves the performances of the technique in the ideal case where the adopted stellar models are in
perfect agreement with real stars and, therefore, sets the absolute minimum variability that is expected to occur only because of the observational uncertainties.

All the recoveries were performed twice. A first set of computations neglects the $\delta \nu$ observational constraint, while a second one adopts it. The comparison of the results under these two different assumptions allowed us to establish the relevance of the small separations in different mass and metallicity ranges, and evolutionary stages.

\subsection{Age estimates}\label{sec.age}

\begin{figure*}
        \centering
        \includegraphics[height=17.8cm,angle=-90]{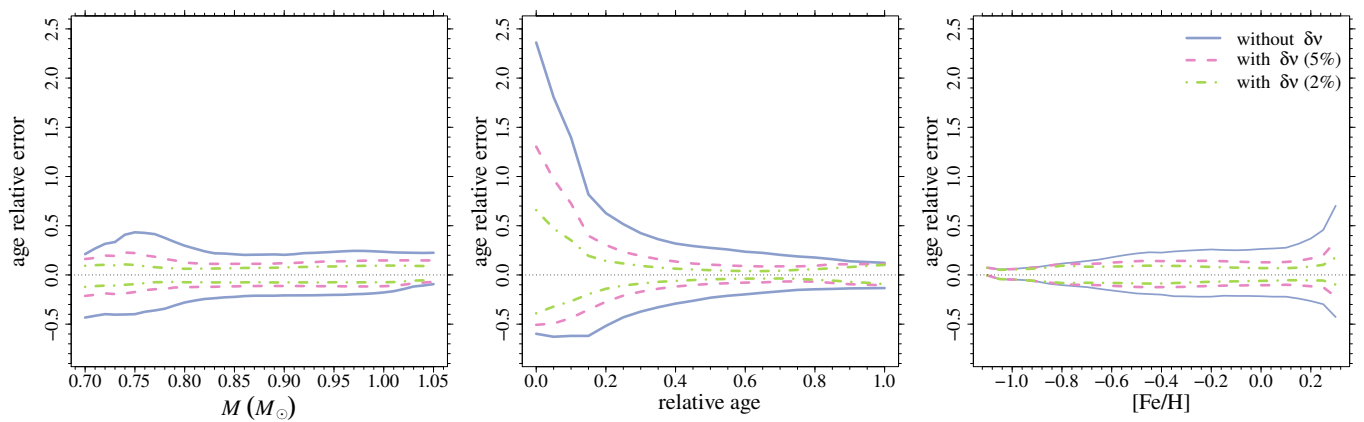}
        \caption{Age-estimate relative errors as a function of the true mass (left panel), 
                relative age (central panel), and metallicity [Fe/H] (right panel) of the
                stars without $\delta \nu$ observational constraint (blue solid line), or adopting $\delta \nu$ with observational errors of 5\% and 2\% (purple dashed line and green dot-dashed line respectively). The lines show the $1 \sigma$ error envelope. A positive relative error indicates that the reconstructed age of the star 
                is overestimated with respect to the true one. }
        \label{fig:age}
\end{figure*}

Figure~\ref{fig:age} shows the dependence of relative errors in
the recovered stellar age as a function of the true stellar mass, the stellar relative age $r$\footnote{The relative age $r$ is defined as the ratio between the age of the star and the age of
the same star at central hydrogen exhaustion. The age is conventionally set to
0 at the ZAMS position.}, and the [Fe/H] value. The [Fe/H] is the current surface value 
for the star, which can be significantly different from the
initial one owing to the microscopic diffusion processes.
The figure shows the relative error envelopes obtained by evaluating
the 16th and 84th quantiles ($1 \sigma$) of the age relative error over a moving window\footnote{The half-width of the window is typically 1/12-1/16 of the range spanned by the independent variable. This choice allows us to maintain the mean relative error in the $1 \sigma$ envelope owing to Monte Carlo sampling at a level of about 1\%, without introducing too much smoothing.}, with a procedure identical to that in \citet{eta}.  The envelope boundaries and the median estimates are also reported in Tables~\ref{tab:amr-vs-M} and \ref{tab:amr-vs-pcage}.

Figure~\ref{fig:age} (scenario with 5\% uncertainty on $\delta \nu$) shows the well known improvement in the age estimates reported in the literature \citep[see e.g.][]{Mathur2012, Lebreton2014, SilvaAguirre2015}. 
The left panel shows the envelope boundary position as a function of the stellar mass. As discussed in \citet{scepter1, eta} edge effects dominate near the grid boundary. As an example, for the lower mass edge ($M = 0.7 M_{\sun}$), no less massive models exist in the grid so a model of this mass can only be confused with a more massive one, which evolve faster. The decrease after about 0.75 $M_{\sun}$ of the position of the upper boundary is instead related to the presence of the hard boundary of 14 Gyr for grid inclusion. While stars more massive than about 0.8 $M_{\sun}$ end their MS life before this limit, this is not the case for less massive stars. For these models, the late MS evolution is neglected, and this exclude the part of their evolution where estimates are the most precise (see central panel in the figure). For $M \geq 0.85 M_{\sun}$ the average envelope half-width without $\delta \nu$ in the observational constraints is about 20\%, while it is as low as 12\% when the value of $\delta \nu$ is available in the fitting. As a comparison \citet{eta} reported about 35\% without the small frequency separation constraint and with a much greater uncertainties in $\Delta \nu$ and $\nu_{\rm max}$ (2.5\% and 5\% respectively). 

\begin{figure}
        \centering
        \includegraphics[height=8.2cm,angle=-90]{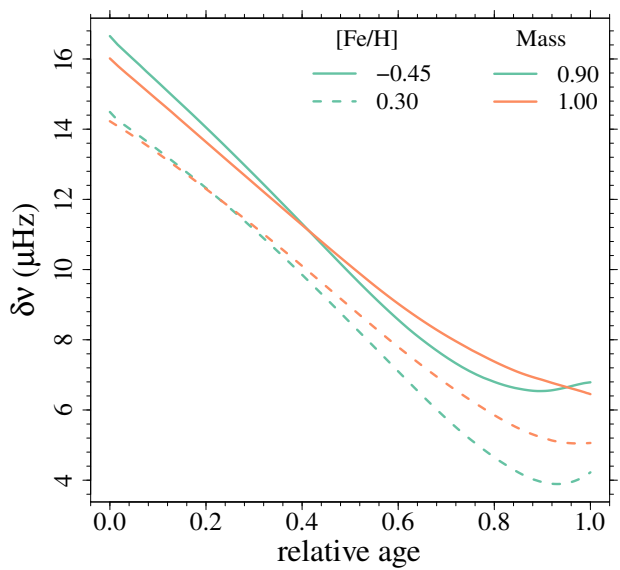}
        \caption{Computed values of the small frequency separations as a function of the relative age of the stars. Curves are labelled according to the considered masses ($M$ = 0.90 and 1.0 $M_{\sun}$) and the metallicity $Z$ of the stars.}
        \label{fig:dnu-r}
\end{figure}

The middle panel shows the performances that can be reached during the MS evolution regardless of the mass. As explained in detail in \citet{eta}, the estimates suffer from a large relative indetermination in the first 20\% of the MS evolution and then shrink to nearly constant, slowly decreasing level. The constant shrinkage is a feature already discussed in literature, and occurs because age estimations are intrinsically easier in
rapidly evolving phases \citep[see e.g. the results in][]{Gai2011, Chaplin2014}.
The mean envelope half-width for $r > 0.3$ is about 22\% without $\delta \nu$ and 11\% with this supplementary constraint. In a departure from \citet{eta}, we do not detect the increase in the envelope width around 80\% of the MS evolution. This is caused by the different mass range considered in the works because the increase in \citet{eta} was caused by the presence of higher mass stars with a convective core. Interestingly, the usefulness of the small frequency separation in improving age estimates vanishes in the last 20\% of the MS. As a reference, the 80\% of the MS evolution roughly corresponds to the turn-off position, with differences due to the mass and the metallicity of the stellar track. Above $r \approx 0.9$ the two envelopes are nearly indistinguishable. This behaviour has a clear explanation behind it, considering the evolution of $\delta \nu$ during the MS evolution. 
Indeed, for stars in the very terminal MS part and in later evolutionary stages, the small separation becomes much less sensitive as an age diagnostic.
As shown in Fig.~\ref{fig:dnu-r}, the small frequency separation decreases during the first 90\% of the MS. In the terminal part of the MS a steady or slightly increasing trend, depending on the mass and the metallicity, appears. Therefore, models in this zone have similar values of $\delta \nu$, explaining its loss of power in core condition predictions.
This is essentially due to the fact that the oscillation modes are not able to grasp the full behaviour of the sound speed gradient in the very deep core of the star at the end of the MS. The modification of the stellar structure, dictated by the increase of the mean molecular weight as a result of the nuclear burning reactions, cannot be effectively probed by $p$ modes.

Indeed, the adoption of the small frequency separation -- or of frequency ratios by \citet{Roxburgh2003}, which are less sensitive to surface corrections -- is known to be beneficial for stars that are less massive than about 1.5 $M_{\sun}$ in the MS evolution. For these stars, the evolutionary tracks in the $\delta \nu$ versus $\Delta \nu$ plain are well-separated.  As
stars evolve off the MS, their tracks converge for the sub-giant and red-giant evolutionary stages \citep[see e.g.][]{White2011,Arentoft2014, Farnir2019}, thus leading to a reduced discriminating potential of the asteroseismic parameters. Further details about the  sensitivity to the frequency ratios and separation on the RGB can be found e.g. in \citet{Montalban2010}.

The right panel in Fig.~\ref{fig:age} shows the envelope boundaries versus [Fe/H]. The trend in the figure stems from the trend in relative age and from
edge effects. In fact, only evolved models, for which the age estimate is more precise, reach surface [Fe/H] below $-0.6$ dex, thanks to the effect of the microscopic diffusion
In contrast, at the upper metallicity edge, only models in their initial evolutionary stages,  young  enough for diffusion to be inefficient, are present. This lead to a larger envelope.

Figure~\ref{fig:age} also shows the results from the scenario that assumes 2\% uncertainty on $\delta \nu$. The same qualitative feature discussed above are conserved, but the error envelope shrinks even further. For the dependence on the stellar mass we obtained an envelope half-width of 8\% for $M \geq 0.85 M_{\sun}$, three quarters than in the case of higher error in $\delta \nu$. Regarding the evolution with the relative age, the mean envelope half-width for relative age above 0.3 is 6\%, about one half than the value reported for the more uncertain scenario.

\begin{figure*}
        \centering
        \includegraphics[height=16.6cm,angle=-90]{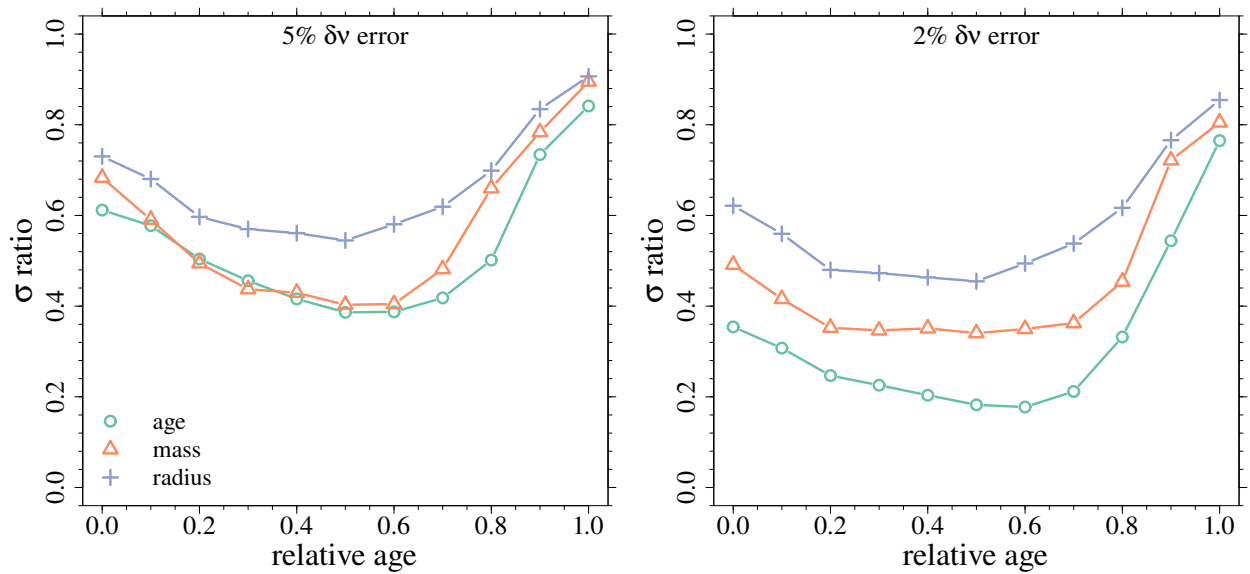}
        \caption{{\it Left}: ratio of the age, mass, and radius error envelope widths between fits obtained with and without $\delta \nu$ as observational constraint, as a function of the stellar relative age. The observational error in $\delta \nu$ is 5\%. {\it Right}: same as in the left panel, but for a 2\% observational error.  }
        \label{fig:relative-sigma}
\end{figure*}

Figure~\ref{fig:relative-sigma} summarises the aforementioned behaviours and shows the ratio of the age error envelope width between fits obtained with and without using $\delta \nu$ as observational constraint as a function of the stellar relative age. The left panel of the figure refers to the scenario that adopts a 5\% error in the small frequency separation, while the right one adopts a 2\% error. In the first case, the ratio of the errors is about 60\% at ZAMS and decreases to a minimum of about 40\% at the middle of the MS evolution. These results agree with the claimed reduction of the error in age by about a factor of two when frequency data are available \citep[e.g.][]{Mathur2012, Lebreton2014, SilvaAguirre2015}.
However, this is only true until about 80\% of the MS evolution because in later stages the error ratio increases again reaching about 80\% in the MS terminal part. 
In the 2\% error scenario the error ratio is about 40\% at ZAMS and decreases to about 20\% in the middle of the MS, a reduction greater than has generally been claimed in the literature. However, even for this scenario, the error ratio climbs again at about 80\% at the end of the MS, so that the average error reduction in the whole MS is about 35\%. 

Overall, the median error in the whole MS for age estimates, marginalised over mass, metallicity and evolutionary phases, are about 15\% and 10\% for observational errors in $\delta \nu$ of 5\% and 2\%, respectively. In the light of these simulations it appears that the most favourable observational targets are stars in the middle of the MS evolution, where the availability of precise and accurate frequency measurements can potentially make the greatest difference.

\subsection{Mass and radius estimates}\label{sec:MR}

\begin{figure*}
        \centering
        \includegraphics[height=16.6cm,angle=-90]{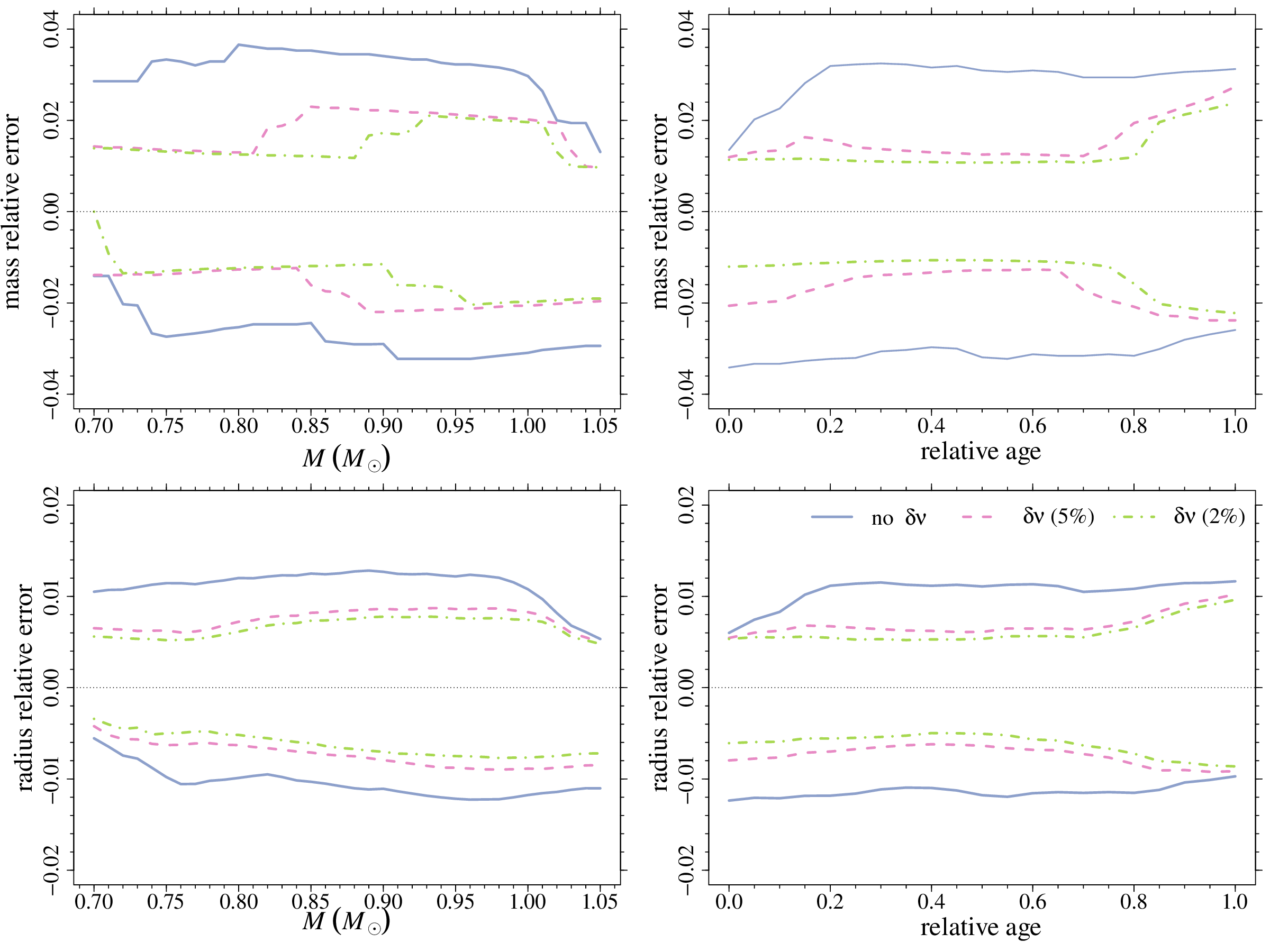}
        \caption{{\it Top row, left}: Mass relative errors as a function of the true mass of the stars, without $\delta \nu$ in the observational constraints (blue solid line), or adopting $\delta \nu$ with observational errors of 5\% and 2\% (purple dashed line and green dot-dashed line). The lines show the $1 \sigma$ error envelope. {\it Right}: same as in the left panel, but as a function of the stellar relative age. {\it Bottom row}: same as in the top row, but for radius estimates.    }
        \label{fig:MR}
\end{figure*}

The contribution of the small separation to the age estimates is clearly shown in Sect.~\ref{sec.age} and its general relevance to the question has already been discussed in the literature \citep[see e.g.][]{Lebreton2009,SilvaAguirre2015, Bellinger2016}. However, the improvement that arises from its availability in the observational constraints is not confined to age. Indeed, it is either mass and radius estimates  that can benefit from it. 

Thanks to the information carried by $\delta \nu$ on the deep layers of the star, the fit has an improved sensitivity to the stellar mass. This effect is  
shown in the top row of Fig.~\ref{fig:MR} that shows the $1 \sigma$ error envelope of the recovered mass as a function of the real mass and the stellar relative age. The figure shows the envelope computed without the $\delta \nu$ observational constraint as well as those that include it with two different error assumptions (namely, 5\% and 2\%). Besides the features dictated by edge effects that have already been extensively discussed in \citet{scepter1}, a clear improvement in the mass estimates is apparent when $\delta \nu$ is available. As has been noted with regard to age estimates, the contribution of $\delta \nu$ tends to vanish in the last 15\% of the evolution (right panel in the figure). As a difference from age estimates, no clear improvement is found adopting a 2\% error in $\delta \nu$ other than in the first 20\% of the MS evolution. 
The bottom row in Fig.~\ref{fig:MR} shows that the same behaviour occurs for radius estimates. 
The position of the $1 \sigma$ envelope boundaries and of the median of the estimates are reported in Tables~\ref{tab:amr-vs-M} and \ref{tab:amr-vs-pcage}.

Figure~\ref{fig:relative-sigma} shows that the improvement provided by the availability of $\delta \nu$ in the fit for mass estimates is nearly identical to that on the age assuming an observational uncertainty of 5\%. An error ratio varying between 60\% and 40\% was found in the first 70\% of the MS evolution, and similar values resulted when assuming an observational uncertainty of 2\%, with the only difference of a slightly decrease of about 10\% in the first 20\% of the MS evolution. The error ratio for radius estimates is somewhat higher, being always over 60\% adopting an observational error of 5\%, and over 50\% with an observational error of 2\%.

The overall error in masses and radii on the whole MS are about 2.2\% and 1.6\% (mass) and 1.0\% and 0.8\% (radius) for observational uncertainties in $\delta \nu$ of 5\% and 2\%, respectively.

\section{Systematic effects}\label{sec:syst}

Section~\ref{sec:results} deals only with random perturbation of observational constraints, which are otherwise supposed to perfectly match the recovery grid. This assumption is clearly overoptimistic when the recovery is applied to real world stars. In this case, systematic discrepancies are expected to occur. It is therefore interesting to explore some systematics that can occur when synthetic objects are sampled from a grid different than the one adopted for the recovery.

Several uncertainty sources can affect the MS evolution of low-mass stars, that is, chemical composition, efficiency of the microscopic diffusion, radiative opacities, nuclear cross section, equation-of-state, mixing scheme, etc. \citep[see e.g.][]{incertezze1, Stancliffe2015}. While it is out of the scope of the present paper to perform a comprehensive analysis of all of these properties, we present in this section a detailed discussion of two of them. This exercise is particularly useful highlighting that the presence of several hidden biases should be carefully considered because the errors from the fitting procedure can't take them in account. As a consequence, the uncertainties in mass, radius and age derived for real stars are probably overly optimistic.

\subsection{Initial helium abundance}\label{sec:dydz}

A first effect that we find is worth exploring is the effect of the choice of initial helium abundance.
In fact, the helium-to-metal enrichment ratio $\Delta Y/\Delta Z$, which is commonly adopted by stellar modellers to select the initial helium
abundance, is quite uncertain \citep{pagel98, jimenez03, gennaro10}. To quantify the impact of this uncertainty,  we
built two synthetic datasets, each of $N = 50\,000$ artificial stars,
by sampling the objects from two non-standard grids with $\Delta Y/\Delta Z$ = 1.0 and 3.0. The
characteristics of the objects was then estimated using the standard grid with $\Delta Y/\Delta Z$ = 2.0 for the recovery. \citet{eta}, in exploring a different mass range and adopted different assumptions on the errors in the asteroseismic quantities, already showed that the variability in the initial helium abundance is negligible for age estimates in the presence of asteroseismic constraints. This occurred owing to compensation effects that are extensively discussed in that paper. 
The change in the initial helium abundances strongly affect, at fixed evolutionary phase, either the age and the effective temperature of an artificial star.  
Thus, a
helium-rich star lies in a zone of the standard grid populated by more massive models, leading to a mass overestimate in the recovery. 
However, helium-rich stars evolve faster than the corresponding standard scenario stars and it so happens that the age bias due to the mass overestimate compensates for the difference in age due to the change in the initial helium. The net effect is a very small age bias. We find a similar behaviour in the present analysis. 

\begin{figure*}
        \centering
        \includegraphics[height=16.6cm,angle=-90]{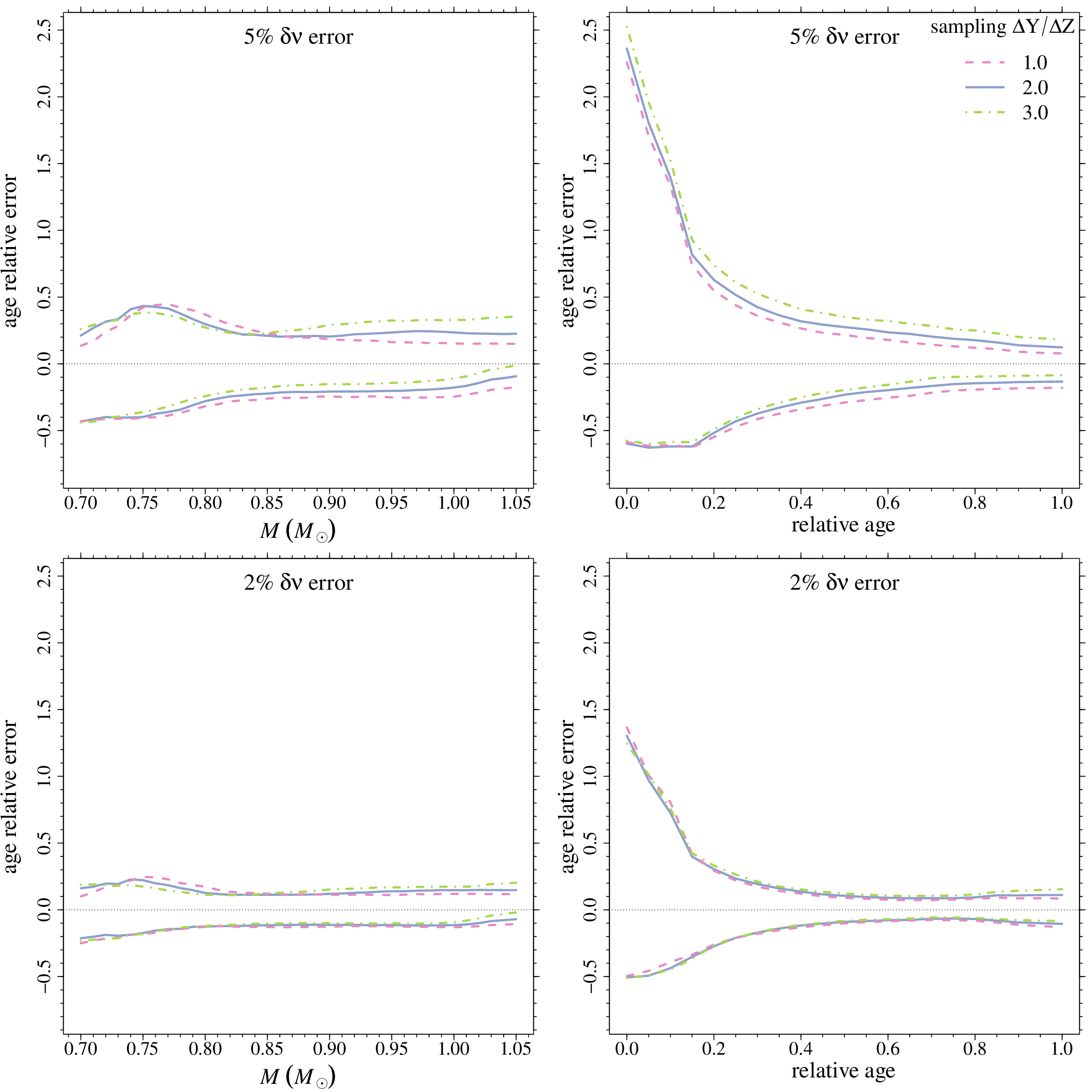}
        \caption{{\it Top row, left}: age-relative error adopting $\delta \nu$ as observational constraints (with observational error of 5\%) versus the true stellar mass. Data were sampled from grid at different $\Delta Y/\Delta Z$  = 1.0, 2.0, and 3.0 and reconstructed over the grid with $\Delta Y/\Delta Z = 2.0$. {\it Right}: same as in the left panel but as a function of the stellar relative age.
        {\it Bottom row}: as in the top row, but using the constraint on $\delta \nu$ with an observational error of 2\%.}
        \label{fig:dydz}
\end{figure*}

Figure~\ref{fig:dydz} and Tables~\ref{tab:ageMR-dydz1} and \ref{tab:ageMR-dydz3} show the position of the age relative error envelope boundaries according to the value of $\Delta Y/\Delta Z$ adopted in the sampling. The top row shows that without the $\delta \nu$ constraint, the age estimates present a marginal variability linked to the uncertainty in the initial helium abundance. The mean offset over the whole MS evolution between the estimates from the $\Delta Y/\Delta Z = 3.0$ and those from $\Delta Y/\Delta Z = 1.0$ is about 10\%.
This variability shrinks to an impressive 1.5\% when the constraint on $\delta \nu$ is added in the fit. Therefore, the availability of the small frequency separation in the fit makes age estimates particularly robust against the indetermination of the initial helium abundance.

It is interesting to note that the same is not true for mass and radius estimates.  Tables~\ref{tab:ageMR-dydz1} and \ref{tab:ageMR-dydz3} show that the fit with the $\delta \nu$ constraint provides masses and radii which are more biased than the ones without this constraint. The difference in the median recovered mass between the $\Delta Y/\Delta Z = 1.0$ and 3.0 is about 1\% without $\delta \nu$, and 2\% with this supplementary constraint. For radius estimates, the differences are about 0.6\% and 0.8\% in the two considered scenarios.
This behaviour stems from the change in the interior layers of the stars -- both on the opacities and in the dimension of the core -- caused by a different assumption on the initial helium abundance. As explained in \citet{scepter1}, the change in the $\Delta Y/\Delta Z$ parameter alters the position of a model on the reconstruction grid, thus leading to biased estimates of both mass and radius. It happens that the further availability of the constraint in $\delta \nu$ worsen the situation because the algorithm, besides matching the effective temperature and the stellar density,  also has to find a structure that provides an adequate small frequency separation. As a consequence, the fit preference is to further bias the mass and radius estimates. Indeed, this behaviour has a simple explanation, and is not linked to the grid of stellar models or the fitting algorithm adopted. The fit was performed on a grid that does not match the one adopted for  synthetic-objects generation. In particular, the age-mass relation is different in these grids. Thus, a better convergence of the fitted solution towards the 'true' value of one of the two of these parameter forces a greater bias in the other. 

The results presented in this section assume the validity of the generally adopted enrichment law that links $Z$ and $Y$ values, with the only freedom of the coefficient residing in the linear relation. Different biases can occur if the metallicity and the initial helium abundance are treated as independent. A full answer to this problem would thus require the computation of stellar models over a dense full grid with varying $Z$ and $Y$. This huge computational effort is, however, outside the aim of this paper. Moreover, most of the databases of stellar tracks that are freely available to the community assume such a dependence between $Z$ and $Y$; therefore, the investigation performed here is particularly relevant because it directly addresses the bias expected in this configuration.

\subsection{Element diffusion}\label{sec:diffusione}

Another source of systematic bias come from the efficiency of the microscopic diffusion. As in \citet{eta}, we quantified a maximum possible bias arising from this factor by adopting an approach slightly different from that adopted in the previous sections.
In this case, we sampled $N = 50\,000$ objects from the standard
grid of models, which takes the element diffusion into account,
and adopted a grid which neglect element diffusion for the recovery.

\begin{figure}
        \centering
        \includegraphics[height=8.4cm,angle=-90]{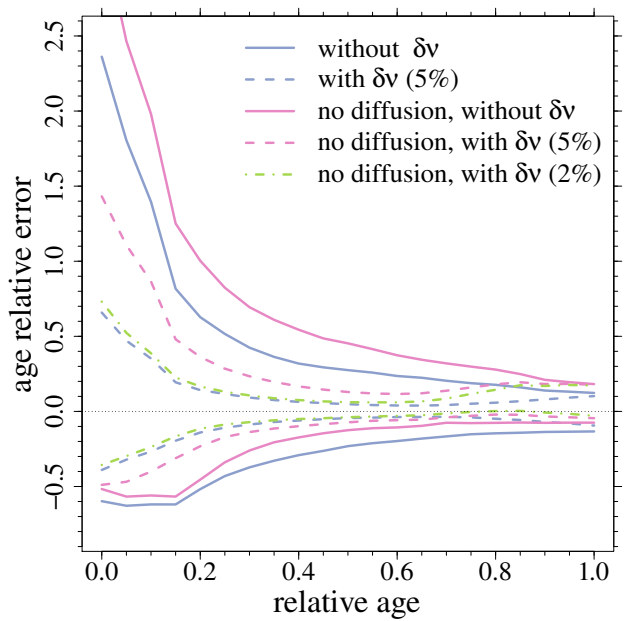}
        \caption{Age-relative error with or without microscopic diffusion, with different assumptions  on observational errors in $\delta \nu$. The blue lines corresponds to the standard recovery and serve as a reference. The other lines show the envelope boundaries when data are sampled from the standard grid and reconstructed on a grid that neglect microscopic diffusion. The magenta solid line corresponds to a recovery without $\delta \nu$; the magenta dashed line corresponds to results obtained with an observational error of 5\% in $\delta \nu$; the green dot-dashed line comes from a scenario with an error of 2\% on  $\delta \nu$.}
        \label{fig:diff}
\end{figure}

The results, presented in Fig.~\ref{fig:diff} and Table~\ref{tab:diffusione}, confirm the findings of \citet{eta}. Without the supplementary constraint provided by $\delta \nu$, the age estimates suffer from a bias of about 20\% in the first part of the MS evolution, which is reduced to about 5\% in the terminal part. The availability of the small frequency separation in the fit strongly reduces these biases, reaching -- for the major part of the MS evolution -- levels below 5\% and 2\% for observational errors of 5\% and 2\%, respectively. For relative age above about 80\%, the relevance of the small frequency separation vanishes, as has been found in findings of the previous sections. 

The estimates of mass and radius show a behaviour similar to that discussed in Sect.~\ref{sec:dydz}, with a greater distortion in the presence of the $\delta \nu$ constraint. However, these biases are so minute -- reaching levels of 1\% level or below -- that they are considered to have no practical relevance.  

As a comparison, we consider the estimation by \citet{Nsamba2018} of  the effect of neglecting the microscopic diffusion on a sample of 34 stars, which obtained median biases of  0.8\%, 2.1\%,
and 16\% in radius, mass, and age, respectively. That research adopted individual frequency in a Bayesian framework and allowed for a supplementary degree of freedom in the value of the mixing-length parameter within the fit. While a direct comparison of the quoted errors is not accurate due to the many differences in the studies, it is interesting to note that the median bias reported in Tab.~\ref{tab:diffusione} for age estimates in the presence of the $\delta \nu$ constraint are significantly lower than that reported by \citet{Nsamba2018}. Indeed, the two quantities do not actually estimate the same effect. While Tab.~\ref{tab:diffusione} presents the theoretical estimates of the diffusion effect in a controlled environment, all the other input being the same between grid and mock data, this is not the case when analysing real data. In the latter case, several uncontrolled discrepancies would exist and their hidden biases would propagate into the final estimates.

\section{Comparison with other pipelines}\label{sec:confronto}

\begin{figure*}
        \centering
        \includegraphics[height=16.4cm,angle=-90]{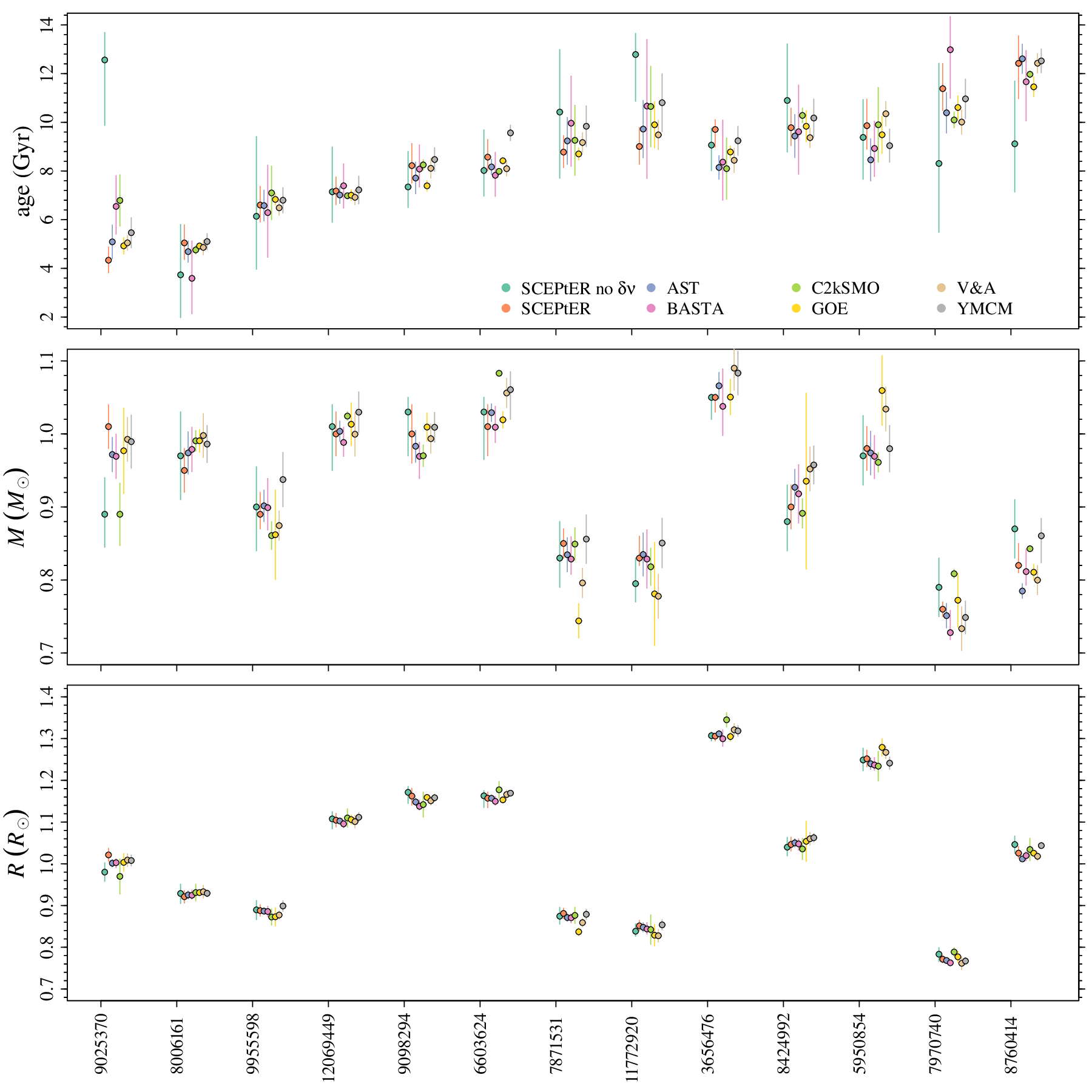}
        \caption{Age, mass, and radius recovered by different pipelines for the thirteen stars selected from the {\it Kepler} asteroseismic LEGACY sample (see text).}
        \label{fig:confronto}
\end{figure*}
 
The results presented in the previous sections shed some light on the minimum biases 
that may arise due to the random observational uncertainties or some systematic effects.
However, the systematic effects considered are only a few of the possible choices made by a modeller when constructing a recovery grid. It is well recognised in the
literature \citep[e.g.][]{Gai2011,Mathur2012,Chaplin2014,SilvaAguirre2017} that the adoption of different grids and estimation procedures can produce estimates with relevant systematic variance.

It is, therefore, interesting to verify how the results discussed so far on the expected variability apply to a real world sample. For this exercise, we selected among the 66 stars in the {\it Kepler} asteroseismic LEGACY sample \citep{Lund2017}, those with expected mass below 1.02 $M_{\sun}$, to avoid as possible edge effects in our estimates. The selection was performed estimating the stellar mass from scaling relations \citep{Ulrich1986, Kjeldsen1995}:
\begin{equation}
\frac{M}{M_{\sun}} = \left( \frac{\nu_{\rm max}}{\nu_{\rm max, \sun}} \right)^3 \left( \frac{\Delta \nu}{\Delta \nu_{\sun}} \right)^{-4} \left( \frac{T_{\rm eff}}{T_{\rm eff, \sun}} \right)^{3/2},
\end{equation}
adopting the \citep{Huber2011}  reference solar values $\Delta \nu_{\sun}$ = 135.1 $\mu$Hz, $\nu_{\rm max, \sun}$ =  3090 $\mu$Hz.
The properties of the 13 selected stars were then estimated adopting the observational constraints provided by \citep{Lund2017}. To avoid possible issues caused by the grid coarseness, we artificially increased the errors in $\Delta \nu$ and $\nu_{\rm max}$ to a minimum level of 0.1\%. Age, mass, and radius were estimated either neglecting the $\delta \nu$ constraint, and taking it into account. The error estimates were obtained by the procedure outlined at the end of Sect.~\ref{sec:fittingML}.

The results of the fit are shown in Fig.~\ref{fig:confronto}. The figure also shows the estimates presented by \citet{SilvaAguirre2017} on the same stars, obtained by six different pipelines (AST, BASTA, C2kSMO, GOE, V\&A, YMCM). The full description of these pipelines is presented in the quoted research. It is worth noting that there is some heterogeneity in the adopted observational constraints and the treatment of surface effects among them. None of these pipelines adopt the same observational constraints or fitting method as the pipeline presented here. Such discrepancies obviously allow for the possibility to quantitatively explore the relevance of systematic errors in the inferred parameters.  

\begin{table*}[ht]
        \caption{Estimated age, mass, and radius of the thirteen stars selected from the LEGACY sample from SCEPtER pipeline.}\label{tab:est-AgeMR}
        \centering
        \begin{tabular}{lrrr}
                \hline\hline
                KIC & Age (Gyr) & Mass ($M_{\sun}$) & Radius ($R_{\sun}$) \\ 
                \hline
                9025370  & 4.34$_{-0.51}^{+0.54}$ & 1.01 $\pm$ 0.03        & 1.02 $\pm$ 0.02 \\ 
                8006161  & 5.05$_{-0.68}^{+0.74}$ & 0.95 $\pm$ 0.03        & 0.92$_{-0.02}^{+0.01}$ \\ 
                9955598  & 6.60$_{-0.69}^{+0.77}$ & 0.89$_{-0.02}^{+0.03}$ & 0.89 $\pm$ 0.01 \\ 
                12069449 & 7.18$_{-0.56}^{+0.57}$ & 1.00 $\pm$ 0.03        & 1.10 $\pm$ 0.02 \\ 
                9098294  & 8.22$_{-0.71}^{+0.91}$ & 1.00 $\pm$ 0.04        & 1.16 $\pm$ 0.02 \\ 
                6603624  & 8.57$_{-0.55}^{+0.72}$ & 1.01$_{-0.04}^{+0.03}$ & 1.16$_{-0.02}^{+0.01}$ \\ 
                7871531  & 8.78$_{-0.63}^{+0.67}$ & 0.85 $\pm$ 0.02        & 0.88 $\pm$ 0.01 \\ 
                11772920 & 9.01$_{-0.73}^{+0.78}$ & 0.83$_{-0.01}^{+0.03}$ & 0.85 $\pm$ 0.01 \\ 
                3656476  & 9.71$_{-0.72}^{+0.40}$ & 1.05$_{-0.02}$         & 1.31 $\pm$ 0.01 \\ 
                8424992  & 9.78$_{-0.74}^{+0.80}$ & 0.90 $\pm$ 0.03        & 1.05 $\pm$ 0.02 \\ 
                5950854  & 9.86$_{-0.96}^{+1.09}$ & 0.98 $\pm$ 0.03        & 1.25 $\pm$ 0.02 \\ 
                7970740  & 11.38$_{-0.89}^{+1.04}$ & 0.76 $\pm$ 0.01       & 0.77 $\pm$ 0.01 \\ 
                8760414  & 12.42$_{-1.45}^{+1.13}$ & 0.82$_{-0.01}^{+0.03}$ & 1.03$_{-0.01}^{+0.02}$  \\ 
                \hline
        \end{tabular}
\end{table*}

Based on Table~\ref{tab:est-AgeMR} and Fig.~\ref{fig:confronto}, it appears that the relevance of the small frequency separation in the fit is evident for few stars. The age, mass, and radius of KIC 9025370, 11772920, 7970740, 8760414 obtained from the SCEPtER pipeline without $\delta \nu$ are very different from the parameters obtained by adding  this observational
constraint into the pipeline.
Besides the estimates obtained by SCEPtER pipeline when neglecting $\delta \nu$, the other pipelines agree quite well, as has been discussed in \citet{SilvaAguirre2017}. The analyses presented in that paper focused on the comparison between the pipelines by adopting the BASTA pipeline as reference. 
Here we supplement  this analysis  by evaluating, on the selected sub-sample, the ratio of the variability caused by the adoption of different pipelines over the average variability, owing to the observational error propagation, on the individual recovered parameters.  

\begin{figure}
        \centering
        \includegraphics[height=8.2cm,angle=-90]{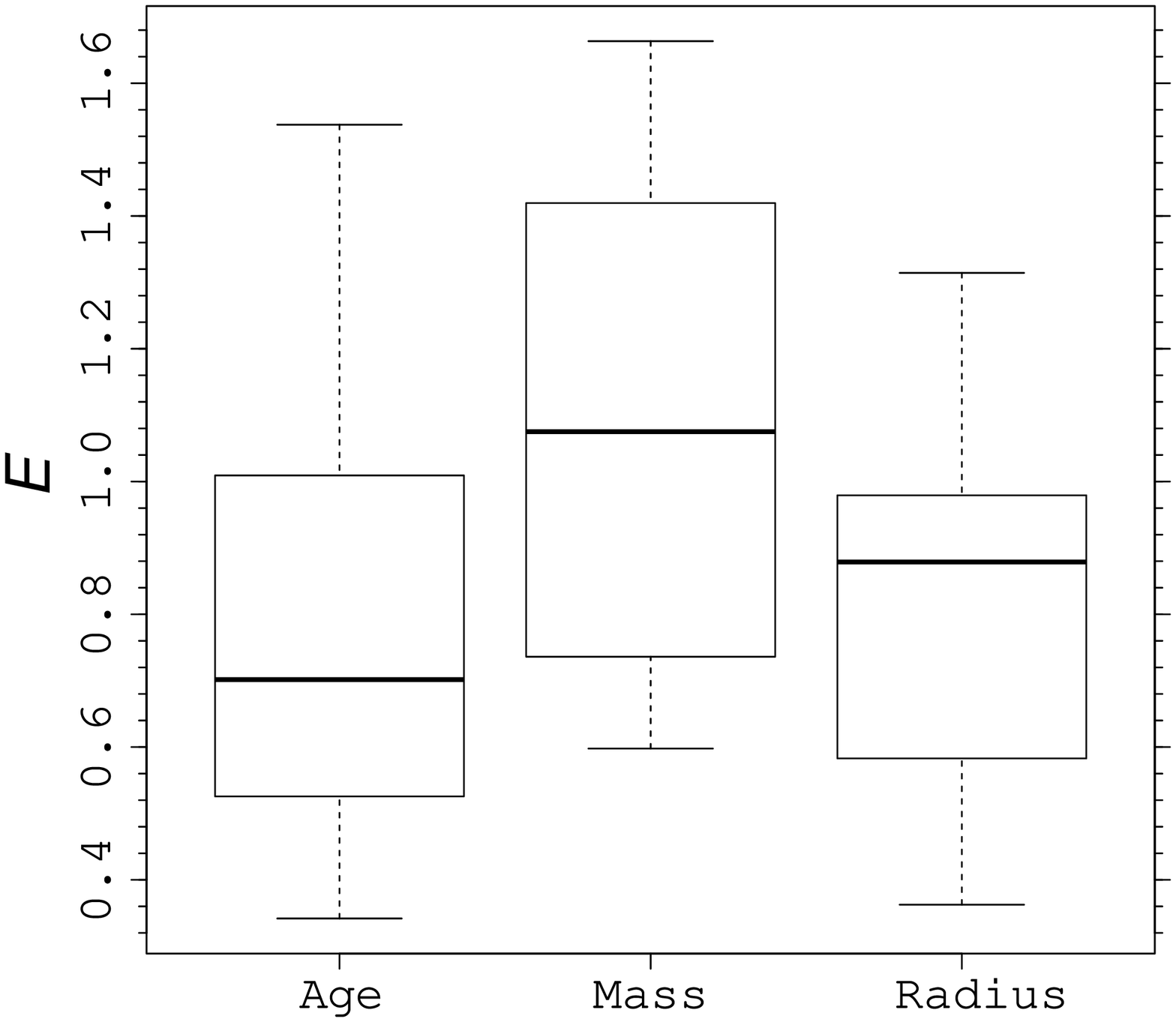}
        \caption{Boxplot of the statistic $E$ comparing the among-pipelines variability with the errors reported by the individual pipelines.  }
        \label{fig:E}
\end{figure}

For this purpose we computed for each star the standard deviation $\sigma_1$ among the estimated values from different pipelines for age, mass, and radius. We also computed the mean value $\sigma_2$ of the errors reported by the different pipelines. To do this, we first computed the quadratic mean of the errors by the individual pipelines (to account for asymmetric errors) and then averaged the values for the different pipelines. Then we compute the ratio of the errors as: 
\begin{equation}
E = \frac{\sigma_1}{\sigma_2}. \label{eq:E}
\end{equation}
A value of $E$ above 1.0 implies that the hidden systematic variability among pipelines is greater than the recovered error in the stellar characteristics.
Figure~\ref{fig:E} shows that the systematic among pipelines has the lowest importance for age estimates, with a median $E \approx 0.7$. Indeed age estimates are affected by the largest errors, thus leading to the lowest intra-pipeline difference. The $E$ values for mass and radius are about 1.1 and 0.9, respectively. Therefore, the detected hidden systematic is nearly as large as the random uncertainties derived by the single pipelines. Ultimately, a conservative, albeit realistic, approach is then to approximately double the error in the recovered stellar characteristics from a single pipeline.  
 
By adding in quadrature $\sigma_1$ and $\sigma_2$ and dividing it by the mean value of the stellar characteristics provided by all the considered pipelines (neglecting the SCEPtER fit without $\delta \nu$), we obtained a precision of about 4.4\%, 1.7\%, and 11\% on the estimated masses, radii, and ages. These values are similar to those reported by \citet{Reese2016} in an hare-and-hounds exercise on 10 MS stars, but adopting different assumptions about the observational uncertainties. The errors obtained in that paper were 3.9\% (mass), 1.5\% (radius), and 23\% (age). The higher quality of data in the LEGACY sample and the different seismic techniques adopted in the researches could explain the difference in the age errors.

\section{Conclusions}\label{sec:conclusions}

We analysed the effect of the availability of the small frequency separation in determining stellar ages, mass, and radius. 
We performed a theoretical investigation aimed to quantify the impact 
of the small frequency separation on the statistical errors and systematic biases affecting stellar age, mass, and radius estimates in a highly controlled framework. 
We performed our investigations on mock datasets of stars sampled from grids of pre-computed stellar models. The approach allows us to establish the minimum possible errors, both statistical and systematic, that affect asteroseismic estimates. We performed the analysis considering stellar models in MS, with masses below 1.05 $M_{\sun}$, allowing us to neglect the uncertainty in the convective core overshooting extension, since these stars burn hydrogen in a radiative core.
 
The study addresses the relevance of the small frequency separation by evaluating the precision and the accuracy of stellar parameters obtained under different assumptions on its observational precision.
In particular we discussed the reconstruction of stellar parameters relying on classical ($T_{\rm eff}$ and [Fe/H]) and global asteroseismic ($\Delta \nu$, $\nu_{\rm max}$, $\delta \nu$) observables of MS field stars. Although several investigations in the literature have addressed similar questions, they are based on a sample of observed stars \citep[see e.g.][]{Bellinger2016, Angelou2017, SilvaAguirre2017, Aerts2018} or they are restricted to specific small samples \citep[e.g.][]{Reese2016}. 

In our study, the parameters of the mock stars were obtained  by the SCEPtER pipeline, a well-tested maximum-likelihood procedure adopted in the past for several investigation on field stars \citep{scepter1, eta, bulge, ML}.
 The analyses were conducted in different configurations. A preliminary investigation was performed sampling the mock data from the same grid adopted for the estimation process. The artificial stars were then subject to random perturbation in the adopted observational range to simulate observational errors. This ideal scenario allowed us to establish the lowest minimum error in the reconstructed stellar age, mass, and radius.
A reference fit was computed, neglecting the presence of $\delta \nu$, which then served as a comparison scenario. Two other fits of the same mock stars were obtained assuming observational errors in $\delta \nu$ of 5\% and 2\% (reachable only for the best targets). 

These investigations highlight the  reported improvement in the age estimates \citep[see e.g.][]{Mathur2012, Lebreton2014, SilvaAguirre2015} in the presence of the $\delta \nu$ constraint. As has already been extensively discussed \citep[see e.g.][]{eta},
the statistical error affecting age estimates is strongly dependent on the stellar evolutionary phase. The error is at its maximum at ZAMS and it stabilises at about 22\% (without $\delta \nu$), 11\% ($\delta \nu$ known at 5\% level), and 6\%  ($\delta \nu$ known at 2\% level) when stars reach 30\% of their evolutionary MS lifetime. This last reduction at about one quarter of what is attainable without $\delta \nu$ in the observational pool is larger than the values -- about one half -- that are commonly reported in the literature. Due to the fact that the present research was performed in a controlled framework, we were able to detect the very best performances achievable. 

Interestingly, this theoretical approach allowed us to put forth evidence that the usefulness of the small frequency separation in improving age estimates vanishes in the last 20\% of the MS, roughly corresponding to the turn-off position. Above the 90\% of the MS evolution, the estimates in the three discussed configurations are indistinguishable. This behaviour is comprehensible considering the evolution of $\delta \nu$ in the last stages of the MS evolution. 
In this evolutionary phase the small frequency separation, which decreased during the first part of the MS evolution shows a steady or slightly increasing trend -- depending on the stellar mass and the metallicity. Therefore, models in the last part of the MS have similar values of $\delta \nu$, with a corresponding loss of power of this constraint in age predictions.

The improvement that arises from $\delta \nu$ availability in the observational constraints is not confined to age.
The availability of $\delta \nu$ in the fit for mass estimates provided an effect nearly identical to that on the age, assuming an observational uncertainty of 5\%. 
This uncertainty allows a reduction of the errors in mass estimates of about 40\% of that obtained without $\delta \nu$. However, in the last 15\% of the MS evolution the impact of the small separation vanishes. A difference with respect to age estimates was that no benefit was detected for mass when adopting an observational error of 2\% in $\delta \nu$. 
The error ratio for radius estimates was a little larger, being always over 60\% adopting an observational error of 5\%, and over 50\% with an observational error of 2\%.

As a major departure from the scenarios discussed here, the adoption of grid based estimates of stellar parameters for real stars is plagued by the presence of unavoidable hidden discrepancies between real world stars and synthetic models. It is therefore relevant to understand as much as possible what biases are expectable when working outside of a controlled framework.
To this aim, we investigated a couple of different scenarios.
In particular, we evaluated the biases arising from the uncertainties in initial helium content (modelled by assuming different values of the helium-to-metal enrichment ratio $\Delta Y/\Delta Z$), and in the microscopic diffusion. These investigations were performed by building several mock catalogues sampled from grids with different assumptions in the initial helium abundances and in the efficiency of the microscopic diffusion, then adopting for the recovery the standard grid.

The estimated age variability atrributed to differences in the initial helium abundance is found to be negligible in the presence of asteroseismic constraints. This occurred owing to compensation effects extensively discussed in \citet{eta}. 
In further detail, we found that in the presence of $\delta \nu,$ the mean difference over the whole MS evolution between the estimates from the $\Delta Y/\Delta Z = 3.0$ and those from $\Delta Y/\Delta Z = 1.0$ shrinks to an impressive 1.5\% (about one seventh of that achievable without $\delta \nu$). Therefore, the availability of the small frequency separation in the fit makes age estimates particularly robust against the indetermination of the initial helium abundance.
It is interesting to note that adopting $\delta \nu$ in the observational pool leads to inferior mass and radius estimates. The difference in the median recovered mass between the $\Delta Y/\Delta Z = 1.0$ and 3.0 is about 1\% without $\delta \nu$, and 2\% with this supplementary constraint. For radius estimates, the differences are about 0.6\% and 0.8\% in the two  scenarios considered.
Indeed, the presence of $\delta \nu$ in the observational pool forces the estimates to match not only the stellar density but also a stellar structure which provides the correct small frequency separation. As a consequence, the fit preference is to further bias the mass and radius estimates.   

With regard to the efficiency of the microscopic diffusion, we investigate the difference in stellar parameters estimates obtained when adopting  a grid that totally neglects the process for
the recovery. 
Without the supplementary constraint provided by $\delta \nu$, the age estimates suffer from a bias of about 20\% in the first part of the MS evolution, which is reduced to about 5\% in the terminal part. The availability of the small frequency separation reduces these biases at levels below 5\% and 2\% for observational errors of 5\% and 2\%, respectively. Also, in this case, the usefulness of the small frequency separation vanishes in the last 20\% of the MS evolution. 
The estimates of mass and radius demonstrate a behaviour that is similar to that which has been discussed for the initial helium abundance, showing a greater distortion in the presence of the $\delta \nu$ constraint. However, these biases are at the level of 1\% or below. 

Given the presence of several assumptions in the link between the seismic indicators and the stellar structure and its evolution (e.g. the link between the small frequency separation and the sound speed gradient; the degeneration in the stellar structure among temperature, chemical composition and sound speed) and that of other unexplored sources of hidden biases (due to the possible difference in the  input physics in the stellar model computations) at present a bias as low as 1\% is probably the very minimum that can be assumed based on the derived stellar parameters. 

The two systematics considered so far cover only a few of the possible choices made by a modeller when constructing a recovery grid. Indeed, it is well-known  \citep[e.g.][]{Gai2011,Mathur2012,Chaplin2014,SilvaAguirre2017} that the adoption of different grids and estimation procedures can produce estimates with relevant systematic variance. To  quantitatively explore how hidden systematics impact on stellar parameter estimates by grid techniques, we applied the SCEPtER pipeline to 13 stars selected from the {\it Kepler} asteroseismic LEGACY sample \citep{Lund2017} with expected masses below 1.02 $M_{\sun}$. A comparison of the results obtained here and in \citet{SilvaAguirre2017} (with six different pipelines) showed a fair agreement of the results, even if the pipelines differ among each other in the adopted observational constraints and in the treatment of surface effects.  
A comparison of the among-pipelines variability with what has been reported individually by each pipeline showed that a conservative but realistic approach is to approximately double the error in the recovered stellar characteristics from a single pipelines.  
This conclusion is possibly overoptimistic because the compared pipelines rely on similar, and in many cases identical, assumptions about the input physics. Larger discrepancies are expected whenever this constraint is relaxed. 
Overall, thanks to the exquisite quality of the LEGACY sample data, we obtained a multi pipeline precision of about 4.4\%, 1.7\%, and 11\% on the estimated masses, radii, and ages.

\begin{acknowledgements}
We thank our anonymous referee for the useful comments and suggestions, that largely improved the paper.     
This work has been supported by PRA Universit\`{a} di Pisa 2018-2019 
(\emph{Le stelle come laboratori cosmici di Fisica fondamentale}, PI: S. Degl'Innocenti) and by INFN (\emph{Iniziativa specifica TAsP}).
\end{acknowledgements}

\bibliographystyle{aa}
\bibliography{biblio}

\appendix

\section{Tables}

In this appendix, we present the tables containing the results quoted throughout the paper.

% latex table generated in R 3.5.1 by xtable 1.8-3 package
% Thu Nov 15 10:22:36 2018
\begin{table*}[ht]
        \caption{Median ($q_{50}$) and $1 \sigma$ envelope boundaries ($q_{16}$ and $q_{84}$) for age, mass, and radius relative errors as a function of the mass of the star. Values are expressed as a percent.} \label{tab:amr-vs-M}
        \centering
        \begin{tabular}{lcccccccc}
                \hline\hline
                & \multicolumn{7}{c}{Mass ($M_{\sun}$)}\\
                quantile & 0.7 & 0.75 & 0.8 & 0.85 & 0.9 & 0.95 & 1.0 & 1.05 \\ 
                \hline
                & \multicolumn{7}{c}{Age}\\
                \hline
                & \multicolumn{7}{c}{without $\delta \nu$}\\
                $q_{16}$ & -43.3 & -39.9 & -28.0 & -22.3 & -20.8 & -20.3 & -17.8 & -9.3 \\ 
                $q_{50}$ & -2.3 & 0.0 & -0.1 & -0.1 & -0.1 & 0.0 & 0.0 & 2.0 \\ 
                $q_{84}$ & 21.2 & 43.3 & 29.7 & 21.0 & 20.4 & 23.5 & 23.5 & 22.5 \\ 
                & \multicolumn{7}{c}{with $\delta \nu$ (5\%)}\\
                $q_{16}$ & -21.3 & -17.4 & -12.6 & -11.6 & -11.3 & -11.4 & -11.5 & -7.0 \\ 
                $q_{50}$ & 0.0 & 0.0 & 0.0 & 0.0 & 0.0 & 0.0 & 0.0 & 0.7 \\ 
                $q_{84}$ & 16.2 & 22.2 & 12.6 & 11.3 & 11.9 & 13.9 & 14.7 & 14.7 \\ 
                & \multicolumn{7}{c}{with $\delta \nu$ (2\%)}\\
                $q_{16}$ & -12.2 & -9.4 & -7.5 & -7.7 & -7.7 & -7.4 & -7.5 & -5.2 \\ 
                $q_{50}$ & 0.0 & 0.0 & 0.0 & 0.0 & 0.0 & 0.0 & 0.0 & 0.2 \\ 
                $q_{84}$ & 9.3 & 10.1 & 6.3 & 6.8 & 7.6 & 8.6 & 9.5 & 9.1 \\ 
                \hline
                & \multicolumn{7}{c}{Mass}\\
                \hline
                & \multicolumn{7}{c}{without $\delta \nu$}\\
                $q_{16}$ & -1.4 & -2.7 & -2.5 & -2.4 & -2.9 & -3.2 & -3.1 & -2.9 \\ 
                $q_{50}$ & 0.0 & 0.0 & 0.0 & 0.0 & 0.0 & 0.0 & 0.0 & 0.0 \\ 
                $q_{84}$ & 2.9 & 3.3 & 3.7 & 3.5 & 3.4 & 3.2 & 3.0 & 1.3 \\ 
                & \multicolumn{7}{c}{with $\delta \nu$ (5\%)}\\
                $q_{16}$ & -1.4 & -1.4 & -1.3 & -1.6 & -2.2 & -2.1 & -2.1 & -2.0 \\ 
                $q_{50}$ & 0.0 & 0.0 & 0.0 & 0.0 & 0.0 & 0.0 & 0.0 & 0.0 \\ 
                $q_{84}$ & 1.4 & 1.4 & 1.3 & 2.3 & 2.2 & 2.1 & 2.0 & 1.0 \\ 
                & \multicolumn{7}{c}{with $\delta \nu$ (2\%)}\\
                $q_{16}$ & 0.0 & -1.3 & -1.2 & -1.2 & -1.1 & -1.8 & -2.0 & -1.9 \\ 
                $q_{50}$ & 0.0 & 0.0 & 0.0 & 0.0 & 0.0 & 0.0 & 0.0 & 0.0 \\ 
                $q_{84}$ & 1.4 & 1.3 & 1.3 & 1.2 & 1.7 & 2.1 & 2.0 & 1.0 \\ 
                \hline
                & \multicolumn{7}{c}{Radius}\\
                \hline
                & \multicolumn{7}{c}{without $\delta \nu$}\\
                $q_{16}$ & -0.6 & -1.0 & -1.0 & -1.0 & -1.1 & -1.2 & -1.2 & -1.1 \\ 
                $q_{50}$ & 0.0 & 0.0 & 0.1 & 0.0 & 0.0 & 0.0 & 0.0 & -0.1 \\ 
                $q_{84}$ & 1.1 & 1.1 & 1.2 & 1.2 & 1.3 & 1.2 & 1.1 & 0.5 \\ 
                & \multicolumn{7}{c}{with $\delta \nu$ (5\%)}\\
                $q_{16}$ & -0.4 & -0.6 & -0.6 & -0.7 & -0.8 & -0.9 & -0.9 & -0.8 \\ 
                $q_{50}$ & 0.0 & 0.0 & 0.0 & 0.0 & 0.0 & 0.0 & 0.0 & -0.1 \\ 
                $q_{84}$ & 0.7 & 0.6 & 0.7 & 0.8 & 0.9 & 0.9 & 0.8 & 0.5 \\ 
                & \multicolumn{7}{c}{with $\delta \nu$ (2\%)}\\
                $q_{16}$ & -0.3 & -0.5 & -0.5 & -0.6 & -0.7 & -0.7 & -0.8 & -0.7 \\ 
                $q_{50}$ & 0.0 & 0.0 & 0.0 & 0.0 & 0.0 & 0.0 & 0.0 & 0.0 \\ 
                $q_{84}$ & 0.6 & 0.5 & 0.6 & 0.7 & 0.8 & 0.8 & 0.7 & 0.5 \\ 
                \hline
        \end{tabular}
        \tablefoot{Typical Monte Carlo relative uncertainty on $q_{16}$ and $q_{84}$ is about
                5\%, while the absolute uncertainty on $q_{50}$ is about 0.2\%.} 
\end{table*}

% latex table generated in R 3.5.1 by xtable 1.8-3 package
% Thu Nov 15 10:22:36 2018
\begin{table*}[ht]
        \caption{Median ($q_{50}$) and $1 \sigma$ envelope boundaries ($q_{16}$ and $q_{84}$) for age, mass, and radius relative errors as a function of relative age of the star. Values are expressed as percent.} \label{tab:amr-vs-pcage}
        \centering
        \begin{tabular}{lrrrrrrrrrrr}
                \hline\hline
                & \multicolumn{11}{c}{Relative age}\\
                quantile & 0.0 & 0.1 & 0.2 & 0.3 & 0.4 & 0.5 & 0.6 & 0.7 & 0.8 & 0.9 & 1.0 \\ 
                \hline
                & \multicolumn{11}{c}{Age}\\
                \hline
                & \multicolumn{11}{c}{without $\delta \nu$}\\
                $q_{16}$ & -59.8 & -61.9 & -51.8 & -37.3 & -29.1 & -23.1 & -19.7 & -16.7 & -14.6 & -13.7 & -13.3 \\ 
                $q_{50}$ & 18.9 & 4.8 & 0.0 & 0.0 & 0.0 & 0.1 & 0.0 & 0.4 & 0.2 & 0.0 & -0.1 \\ 
                $q_{84}$ & 236.2 & 139.5 & 62.8 & 42.4 & 31.9 & 27.5 & 23.6 & 20.4 & 17.7 & 13.9 & 12.3 \\ 
                & \multicolumn{11}{c}{with $\delta \nu$ (5\%)}\\
                $q_{16}$ & -50.7 & -43.8 & -27.3 & -17.0 & -11.8 & -9.1 & -7.8 & -6.7 & -6.9 & -9.5 & -10.5 \\ 
                $q_{50}$ & 4.2 & 1.7 & 0.0 & 0.0 & 0.0 & 0.0 & 0.0 & 0.0 & 0.0 & 0.0 & 0.0 \\ 
                $q_{84}$ & 130.4 & 72.6 & 30.4 & 19.3 & 13.6 & 10.5 & 9.0 & 8.7 & 9.3 & 10.8 & 11.1 \\ 
                & \multicolumn{11}{c}{with $\delta \nu$ (2\%)}\\
                $q_{16}$ & -39.0 & -26.7 & -14.1 & -8.8 & -6.1 & -4.4 & -3.8 & -3.6 & -4.9 & -7.0 & -9.4 \\ 
                $q_{50}$ & 0.0 & 0.0 & 0.0 & 0.0 & 0.0 & 0.0 & 0.0 & 0.0 & 0.0 & 0.0 & 0.0 \\ 
                $q_{84}$ & 65.8 & 35.2 & 14.2 & 9.2 & 6.3 & 4.8 & 3.9 & 4.2 & 5.8 & 8.0 & 10.2 \\
                \hline
                & \multicolumn{11}{c}{Mass}\\
                \hline
                & \multicolumn{11}{c}{without $\delta \nu$}\\
                $q_{16}$ & -3.4 & -3.3 & -3.2 & -3.1 & -3.0 & -3.2 & -3.1 & -3.2 & -3.2 & -2.8 & -2.6 \\ 
                $q_{50}$ & 0.0 & 0.0 & 0.0 & 0.0 & 0.0 & 0.0 & 0.0 & 0.0 & 0.0 & 0.0 & 0.0 \\ 
                $q_{84}$ & 1.4 & 2.3 & 3.2 & 3.2 & 3.2 & 3.1 & 3.1 & 2.9 & 2.9 & 3.1 & 3.1 \\ 
                & \multicolumn{11}{c}{with $\delta \nu$ (5\%)}\\
                $q_{16}$ & -2.1 & -2.0 & -1.6 & -1.4 & -1.3 & -1.3 & -1.3 & -1.7 & -2.1 & -2.3 & -2.4 \\ 
                $q_{50}$ & 0.0 & 0.0 & 0.0 & 0.0 & 0.0 & 0.0 & 0.0 & 0.0 & 0.0 & 0.0 & 0.0 \\ 
                $q_{84}$ & 1.2 & 1.3 & 1.6 & 1.4 & 1.3 & 1.3 & 1.3 & 1.2 & 1.9 & 2.3 & 2.7 \\ 
                & \multicolumn{11}{c}{with $\delta \nu$ (2\%)}\\
                $q_{16}$ & -1.2 & -1.2 & -1.1 & -1.1 & -1.1 & -1.1 & -1.1 & -1.1 & -1.6 & -2.1 & -2.2 \\ 
                $q_{50}$ & 0.0 & 0.0 & 0.0 & 0.0 & 0.0 & 0.0 & 0.0 & 0.0 & 0.0 & 0.0 & 0.0 \\ 
                $q_{84}$ & 1.1 & 1.1 & 1.1 & 1.1 & 1.1 & 1.1 & 1.1 & 1.1 & 1.2 & 2.1 & 2.4 \\ 
                \hline
                & \multicolumn{11}{c}{Radius}\\
                \hline
                & \multicolumn{11}{c}{without $\delta \nu$}\\
                $q_{16}$ & -1.2 & -1.2 & -1.2 & -1.1 & -1.1 & -1.2 & -1.2 & -1.2 & -1.2 & -1.0 & -1.0 \\ 
                $q_{50}$ & -0.2 & -0.1 & 0.0 & 0.0 & 0.0 & 0.0 & 0.0 & 0.0 & 0.0 & 0.0 & 0.0 \\ 
                $q_{84}$ & 0.6 & 0.8 & 1.1 & 1.2 & 1.1 & 1.1 & 1.1 & 1.0 & 1.1 & 1.1 & 1.2 \\ 
                & \multicolumn{11}{c}{with $\delta \nu$ (5\%)}\\
                $q_{16}$ & -0.8 & -0.8 & -0.7 & -0.7 & -0.6 & -0.6 & -0.7 & -0.7 & -0.8 & -0.9 & -0.9 \\ 
                $q_{50}$ & -0.1 & 0.0 & 0.0 & 0.0 & 0.0 & 0.0 & 0.0 & 0.0 & 0.0 & 0.0 & 0.0 \\ 
                $q_{84}$ & 0.5 & 0.6 & 0.7 & 0.6 & 0.6 & 0.6 & 0.6 & 0.6 & 0.7 & 0.9 & 1.0 \\ 
                & \multicolumn{11}{c}{with $\delta \nu$ (2\%)}\\
                $q_{16}$ & -0.6 & -0.6 & -0.6 & -0.5 & -0.5 & -0.5 & -0.6 & -0.6 & -0.7 & -0.8 & -0.9 \\ 
                $q_{50}$ & 0.0 & 0.0 & 0.0 & 0.0 & 0.0 & 0.0 & 0.0 & 0.0 & 0.0 & 0.0 & 0.0 \\ 
                $q_{84}$ & 0.5 & 0.5 & 0.5 & 0.5 & 0.5 & 0.5 & 0.6 & 0.6 & 0.7 & 0.9 & 1.0 \\
                \hline
        \end{tabular}
\end{table*}

\begin{table*}[ht]
        \caption{As in Table~\ref{tab:amr-vs-pcage}, but for mock data sampled from a grid with $\Delta Y/\Delta Z = 1.0$.} \label{tab:ageMR-dydz1} 
        \centering
        \begin{tabular}{lrrrrrrrrrrr}
                \hline\hline
                & \multicolumn{11}{c}{Relative age}\\
                quantile & 0.0 & 0.1 & 0.2 & 0.3 & 0.4 & 0.5 & 0.6 & 0.7 & 0.8 & 0.9 & 1.0 \\ 
                \hline
                & \multicolumn{11}{c}{Age}\\
                \hline
                & \multicolumn{11}{c}{without $\delta \nu$}\\
                $q_{16}$ & -58.7 & -60.8 & -54.8 & -41.4 & -34.0 & -29.0 & -25.3 & -21.7 & -19.3 & -18.4 & -18.0 \\ 
                $q_{50}$ & 24.6 & 4.2 & -4.4 & -5.6 & -5.4 & -4.6 & -4.4 & -3.9 & -4.1 & -4.6 & -4.8 \\ 
                $q_{84}$ & 225.8 & 133.6 & 54.8 & 35.9 & 26.5 & 21.7 & 18.0 & 14.2 & 12.0 & 9.0 & 7.8 \\ 
                & \multicolumn{11}{c}{with $\delta \nu$ (5\%)}\\
                $q_{16}$ & -49.6 & -39.2 & -26.0 & -18.0 & -13.0 & -10.2 & -8.6 & -7.6 & -8.2 & -11.3 & -12.9 \\ 
                $q_{50}$ & 16.9 & 5.6 & -0.4 & -0.8 & -0.8 & -0.8 & -0.7 & -0.3 & -0.3 & -1.5 & -2.4 \\ 
                $q_{84}$ & 136.6 & 80.6 & 29.3 & 17.3 & 12.1 & 9.0 & 7.5 & 7.3 & 8.3 & 8.6 & 8.4 \\ 
                \hline
                & \multicolumn{11}{c}{Mass}\\
                \hline
                & \multicolumn{11}{c}{without $\delta \nu$}\\
                $q_{16}$ & -4.4 & -4.2 & -4.0 & -4.0 & -4.0 & -4.0 & -3.9 & -3.9 & -3.8 & -3.4 & -3.3 \\ 
                $q_{50}$ & -1.3 & -1.2 & -1.0 & -0.9 & -0.6 & -0.6 & -0.7 & -0.9 & -0.5 & -0.4 & -0.3 \\ 
                $q_{84}$ & 0.6 & 1.3 & 2.3 & 2.6 & 2.6 & 2.7 & 2.6 & 2.4 & 2.5 & 2.7 & 2.8 \\ 
                & \multicolumn{11}{c}{with $\delta \nu$ (5\%)}\\
                $q_{16}$ & -3.3 & -3.2 & -2.9 & -2.9 & -2.8 & -2.9 & -3.0 & -3.3 & -3.8 & -3.8 & -3.6 \\ 
                $q_{50}$ & -1.2 & -1.2 & -1.1 & -1.2 & -1.2 & -1.2 & -1.2 & -1.2 & -1.2 & -1.0 & -0.8 \\ 
                $q_{84}$ & 0.0 & 0.4 & 0.7 & 0.5 & 0.0 & 0.0 & 0.0 & 0.0 & 1.0 & 1.7 & 2.1 \\ 
                \hline
                & \multicolumn{11}{c}{Radius}\\
                \hline
                & \multicolumn{11}{c}{without $\delta \nu$}\\
                $q_{16}$ & -1.7 & -1.6 & -1.5 & -1.5 & -1.5 & -1.5 & -1.5 & -1.5 & -1.5 & -1.3 & -1.2 \\ 
                $q_{50}$ & -0.6 & -0.5 & -0.4 & -0.3 & -0.3 & -0.3 & -0.3 & -0.3 & -0.3 & -0.2 & -0.2 \\ 
                $q_{84}$ & 0.2 & 0.5 & 0.8 & 0.9 & 0.9 & 0.9 & 0.9 & 0.8 & 0.9 & 1.0 & 1.0 \\ 
                & \multicolumn{11}{c}{with $\delta \nu$ (5\%)}\\
                $q_{16}$ & -1.3 & -1.3 & -1.2 & -1.2 & -1.2 & -1.2 & -1.2 & -1.3 & -1.5 & -1.4 & -1.4 \\ 
                $q_{50}$ & -0.5 & -0.5 & -0.4 & -0.5 & -0.5 & -0.5 & -0.5 & -0.5 & -0.5 & -0.4 & -0.3 \\ 
                $q_{84}$ & 0.2 & 0.2 & 0.3 & 0.3 & 0.2 & 0.2 & 0.2 & 0.3 & 0.4 & 0.6 & 0.8 \\ 
                \hline
        \end{tabular}
\end{table*}

\begin{table*}[ht]
        \caption{As in Table~\ref{tab:amr-vs-pcage}, but for mock data sampled from a grid with $\Delta Y/\Delta Z = 3.0$.} \label{tab:ageMR-dydz3} 
        \centering
        \begin{tabular}{lrrrrrrrrrrr}
                \hline\hline
                & \multicolumn{11}{c}{Relative age}\\
                quantile & 0.0 & 0.1 & 0.2 & 0.3 & 0.4 & 0.5 & 0.6 & 0.7 & 0.8 & 0.9 & 1.0 \\ 
                \hline
                & \multicolumn{11}{c}{Age}\\
                \hline
                & \multicolumn{11}{c}{without $\delta \nu$}\\
                $q_{16}$ & -57.6 & -58.5 & -49.1 & -34.0 & -25.2 & -19.7 & -15.7 & -10.7 & -9.6 & -9.0 & -8.5 \\ 
                $q_{50}$ & 28.7 & 13.8 & 6.8 & 6.2 & 5.5 & 5.4 & 5.9 & 7.0 & 5.6 & 4.4 & 4.1 \\ 
                $q_{84}$ & 252.4 & 152.6 & 73.9 & 52.5 & 40.9 & 35.1 & 32.0 & 28.4 & 24.9 & 20.0 & 18.3 \\ 
                & \multicolumn{11}{c}{with $\delta \nu$ (5\%)}\\
                $q_{16}$ & -51.0 & -44.7 & -26.7 & -16.8 & -11.3 & -8.2 & -7.0 & -5.7 & -6.1 & -7.6 & -8.4 \\ 
                $q_{50}$ & 8.8 & 5.0 & 1.8 & 1.8 & 1.8 & 1.5 & 1.2 & 1.8 & 1.9 & 2.3 & 2.8 \\ 
                $q_{84}$& 124.4 & 75.7 & 33.1 & 21.2 & 15.2 & 12.3 & 10.7 & 10.6 & 11.5 & 14.3 & 15.4 \\ 
                \hline
                & \multicolumn{11}{c}{Mass}\\
                \hline
                & \multicolumn{11}{c}{without $\delta \nu$}\\
                $q_{16}$ & -2.4 & -2.5 & -2.6 & -2.5 & -2.5 & -2.6 & -2.5 & -2.6 & -2.6 & -2.4 & -2.2 \\ 
                $q_{50}$ & 0.2 & 0.6 & 0.7 & 0.8 & 0.7 & 0.6 & 0.5 & 0.3 & 0.5 & 0.5 & 0.6 \\ 
                $q_{84}$ & 2.5 & 3.1 & 4.0 & 4.1 & 4.1 & 4.0 & 3.9 & 3.4 & 3.5 & 3.5 & 3.5 \\ 
                & \multicolumn{11}{c}{with $\delta \nu$ (5\%)}\\
                $q_{16}$ & -1.1 & -1.1 & -1.0 & -0.5 & 0.0 & 0.0 & 0.0 & -0.5 & -1.0 & -1.7 & -2.0 \\ 
                $q_{50}$ & 0.7 & 1.0 & 1.1 & 1.1 & 1.2 & 1.2 & 1.2 & 1.1 & 1.1 & 1.0 & 1.0 \\ 
                $q_{84}$ & 2.3 & 2.5 & 2.7 & 2.7 & 2.7 & 2.7 & 2.7 & 3.0 & 3.3 & 3.5 & 3.7 \\ 
                \hline
                & \multicolumn{11}{c}{Radius}\\
                \hline
                & \multicolumn{11}{c}{without $\delta \nu$}\\
                $q_{16}$ & -0.8 & -0.9 & -0.9 & -0.9 & -0.8 & -0.9 & -0.9 & -0.9 & -0.9 & -0.8 & -0.8 \\ 
                $q_{50}$ & 0.2 & 0.2 & 0.3 & 0.3 & 0.3 & 0.3 & 0.2 & 0.2 & 0.2 & 0.2 & 0.2 \\ 
                $q_{84}$ & 1.0 & 1.2 & 1.5 & 1.5 & 1.5 & 1.5 & 1.4 & 1.3 & 1.3 & 1.3 & 1.3 \\ 
                & \multicolumn{11}{c}{with $\delta \nu$ (5\%)}\\
                $q_{16}$ & -0.4 & -0.4 & -0.3 & -0.3 & -0.3 & -0.3 & -0.2 & -0.3 & -0.4 & -0.6 & -0.7 \\ 
                $q_{50}$ & 0.3 & 0.3 & 0.4 & 0.4 & 0.4 & 0.4 & 0.5 & 0.4 & 0.4 & 0.4 & 0.3 \\ 
                $q_{84}$ & 1.0 & 1.1 & 1.1 & 1.1 & 1.2 & 1.1 & 1.1 & 1.2 & 1.3 & 1.3 & 1.4 \\ 
                \hline
        \end{tabular}
\end{table*}

\begin{table*}[ht]
        \caption{As in Table~\ref{tab:amr-vs-pcage}, but for mock data sampled from a grid with microscopic diffusion and reconstructed on a grid where it is neglected (see also Fig.~\ref{fig:diff}).} \label{tab:diffusione} 
        \centering
        \begin{tabular}{lrrrrrrrrrrr}
                \hline\hline
                & \multicolumn{11}{c}{Relative age}\\
                quantile & 0.0 & 0.1 & 0.2 & 0.3 & 0.4 & 0.5 & 0.6 & 0.7 & 0.8 & 0.9 & 1.0 \\ 
                \hline
                & \multicolumn{11}{c}{Age}\\
                \hline
                & \multicolumn{11}{c}{without $\delta \nu$}\\
                $q_{16}$ & -51.7 & -56.0 & -45.4 & -26.0 & -17.3 & -12.5 & -10.7 & -7.6 & -7.6 & -7.6 & -7.5 \\ 
                $q_{50}$ & 42.4 & 30.0 & 23.5 & 19.9 & 15.7 & 13.1 & 10.7 & 9.5 & 8.0 & 5.4 & 4.3 \\ 
                $q_{84}$ & 321.9 & 198.0 & 100.3 & 69.4 & 54.4 & 45.3 & 37.4 & 32.0 & 27.8 & 21.0 & 18.2 \\ 
                & \multicolumn{11}{c}{with $\delta \nu$ (5\%)}\\
                $q_{16}$ & -48.9 & -40.1 & -23.2 & -13.8 & -9.8 & -7.3 & -5.9 & -3.9 & -2.1 & -3.4 & -4.5 \\ 
                $q_{50}$ & 17.1 & 10.0 & 6.0 & 3.8 & 2.4 & 2.0 & 2.0 & 3.3 & 6.1 & 6.0 & 5.3 \\ 
                $q_{84}$ & 143.2 & 86.5 & 36.3 & 23.4 & 16.7 & 12.9 & 11.6 & 13.8 & 18.0 & 18.3 & 17.8 \\
                & \multicolumn{11}{c}{with $\delta \nu$ (2\%)}\\ 
                $q_{16}$ & -35.6 & -23.7 & -11.5 & -7.1 & -5.1 & -3.9 & -3.0 & -1.3 & 0.3 & -0.8 & -3.0 \\ 
                $q_{50}$ & 5.8 & 4.1 & 2.0 & 1.2 & 0.7 & 0.4 & 0.7 & 2.4 & 5.9 & 6.5 & 6.1 \\ 
                $q_{84}$ & 73.1 & 38.7 & 16.6 & 10.7 & 7.5 & 6.1 & 6.0 & 8.5 & 14.4 & 16.9 & 17.2 \\ 
                \hline
                & \multicolumn{11}{c}{Mass}\\
                \hline
                & \multicolumn{11}{c}{without $\delta \nu$}\\
                $q_{16}$ & -3.8 & -3.7 & -3.7 & -3.4 & -3.3 & -3.5 & -3.6 & -3.8 & -3.8 & -3.4 & -3.2 \\ 
                $q_{50}$ & -0.3 & 0.0 & 0.0 & 0.0 & 0.0 & 0.0 & 0.0 & -0.6 & -0.6 & -0.4 & 0.0 \\ 
                $q_{84}$ & 1.4 & 2.2 & 3.2 & 3.0 & 2.9 & 2.8 & 2.7 & 2.4 & 2.4 & 2.5 & 2.5 \\ 
                & \multicolumn{11}{c}{with $\delta \nu$ (5\%)}\\
                $q_{16}$ & -1.7 & -1.4 & -1.1 & -1.0 & -0.5 & -0.5 & -1.0 & -2.0 & -2.9 & -3.0 & -2.9 \\ 
                $q_{50}$ & 0.0 & 0.0 & 0.5 & 1.0 & 1.1 & 1.1 & 1.1 & 0.0 & 0.0 & 0.0 & 0.0 \\ 
                $q_{84}$ & 1.3 & 1.6 & 2.3 & 2.5 & 2.6 & 2.6 & 2.5 & 2.2 & 2.0 & 2.2 & 2.3 \\ 
                & \multicolumn{11}{c}{with $\delta \nu$ (2\%)}\\
                $q_{16}$ & -1.1 & -1.0 & 0.0 & 0.0 & 0.0 & 0.0 & 0.0 & -1.1 & -2.4 & -2.5 & -2.2 \\ 
                $q_{50}$ & 0.0 & 0.0 & 1.0 & 1.1 & 1.2 & 1.2 & 1.1 & 0.7 & 0.0 & 0.0 & 0.0 \\ 
                $q_{84}$ & 1.2 & 1.3 & 1.6 & 2.3 & 2.5 & 2.5 & 2.4 & 2.2 & 1.5 & 2.2 & 2.4 \\ 
                \hline
                & \multicolumn{11}{c}{Radius}\\
                \hline
                & \multicolumn{11}{c}{without $\delta \nu$}\\
                $q_{16}$ & -1.4 & -1.4 & -1.4 & -1.3 & -1.3 & -1.4 & -1.4 & -1.6 & -1.6 & -1.5 & -1.4 \\ 
                $q_{50}$ & -0.2 & -0.2 & -0.2 & -0.2 & -0.1 & -0.2 & -0.2 & -0.4 & -0.4 & -0.3 & -0.2 \\ 
                $q_{84}$ & 0.6 & 0.8 & 1.1 & 1.0 & 1.0 & 0.9 & 0.9 & 0.8 & 0.8 & 0.8 & 0.9 \\ 
                & \multicolumn{11}{c}{with $\delta \nu$ (5\%)}\\
                $q_{16}$ & -0.7 & -0.7 & -0.5 & -0.4 & -0.4 & -0.4 & -0.6 & -0.9 & -1.3 & -1.3 & -1.2 \\ 
                $q_{50}$ & -0.1 & 0.0 & 0.1 & 0.2 & 0.3 & 0.3 & 0.2 & 0.0 & -0.2 & -0.2 & -0.2 \\ 
                $q_{84}$ & 0.6 & 0.7 & 0.8 & 0.9 & 0.9 & 0.9 & 0.9 & 0.8 & 0.7 & 0.7 & 0.8 \\ 
                & \multicolumn{11}{c}{with $\delta \nu$ (2\%)}\\
                $q_{16}$ & -0.5 & -0.4 & -0.3 & -0.3 & -0.3 & -0.3 & -0.4 & -0.8 & -1.2 & -1.1 & -1.0 \\ 
                $q_{50}$ & -0.0 & 0.1 & 0.2 & 0.3 & 0.3 & 0.3 & 0.3 & 0.1 & -0.2 & -0.1 & -0.0 \\ 
                $q_{84}$ & 0.6 & 0.6 & 0.7 & 0.8 & 0.9 & 0.9 & 0.9 & 0.7 & 0.6 & 0.8 & 0.9 \\ 
                \hline
        \end{tabular}
\end{table*}

\end{document}